\documentstyle{article}

\def\tilde{\widetilde}
\def\bar{\overline}
\def\hat{\widehat}
\def\*{\star}
\def\({\left(}          
\def\){\right)}         
\def\[{\left[}          
\def\]{\right]}

\def\frac#1#2{{#1 \over #2}}            
\def\inv#1{{1 \over #1}}

\def\d{\partial}

\def\vev#1{\langle #1 \rangle}

\def\2pi{\hbox{$2\pi i$}}

\def\dsl{\raise.15ex\hbox{/}\kern-.57em\partial}
\def\Dsl{\,\raise.15ex\hbox{/}\mkern-.13.5mu D}
\def\ga{\gamma}         
\def\b{\beta}
\def\al{\alpha}

\def\la{\lambda}        
         
\def\om{\omega}         
\def\sig{\sigma}        

              \def\CL{{\cal L}}
              \def\CO{{\cal O}}
              
       \def\CT{{\cal T}}

\def\debut{ \begin{eqnarray} }
\def\fin{ \end{eqnarray} }
\def\non{ \nonumber }

\def\presentation{
\voffset -.50in
\hoffset -.19in
\oddsidemargin 0in \evensidemargin 0in
\marginparwidth .75in \marginparsep 7pt \topmargin 0in
\headheight 12pt \headsep .25in
\footheight 18pt \footskip .35in
\textheight 9.5in \textwidth 6.5in
\columnsep 10pt \columnseprule 0pt }

\presentation

\begin{document}

\rightline{SPhT-96-018}
\rightline{LPTHE-96-09}
\vskip 1cm
\centerline{\LARGE Quantization of Solitons and}
\bigskip
\centerline{\LARGE the Restricted Sine-Gordon Model.}
\vskip 2cm
\centerline{\large O. Babelon ${}^{a\ 0}$, D. Bernard ${}^b$
\footnote[0]{Membre du CNRS} and F.A. Smirnov ${}^a$
\footnote[1]{On leave from Steklov Mathematical Institute,
Fontanka 27, St. Petersburg, 191011, Russia} }
\vskip1cm
\centerline{ ${}^a$ Laboratoire de Physique Th\'eorique et Hautes
Energies \footnote[2]{\it Laboratoire associ\'e au CNRS.}}
\centerline{ Universit\'e Pierre et Marie Curie, Tour 16 1$^{er}$
		\'etage, 4 place Jussieu}
\centerline{75252 Paris cedex 05-France}
\bigskip
 \centerline{ ${}^b$ Service de Physique Th\'eorique de Saclay
\footnote[3]{\it Laboratoire de la Direction des Sciences de la
Mati\`ere du Commissariat \`a l'Energie Atomique.}}
\centerline{F-91191, Gif-sur-Yvette, France.}
 \vskip2cm
{\bf Abstract.} We show how to compute form factors, matrix elements
of local fields, in the restricted
sine-Gordon model, at the reflectionless points, by quantizing solitons.
 We introduce (quantum) separated variables in which the
Hamiltonians are expressed in terms of (quantum)
$\tau$-functions. We
explicitly describe the soliton wave functions, and we explain how the
restriction is related to an unusual hermitian structure. We also present
a semi-classical analysis which enlightens the fact that the restricted
sine-Gordon model corresponds to an analytical continuation of the
sine-Gordon model, intermediate between sine-Gordon and KdV. \\
\newpage

\section{Introduction}

About 20 years ago the work on quantization
of integrable models of Quantum Field Theory
started with the idea of quantizing the classical soliton solutions
\cite{dhn,fk}. Important results were achieved in this way,
in particular for the sine-Gordon (SG) theory, the semi-classical
spectrum of excitations (which happens to be exact quantum-mechanically)
and the semi-classical approximation for the soliton S-matrix
were found. The semi-classical S-matrix allowed to
guess the exact S-matrix in the reflectionless case \cite{fk},
and this was used further as a fundamental input in the bootstrap
construction of the S-matrix for arbitrary coupling \cite{zz}.
Later, however the idea of direct quantization of solitons
was abandoned in favor of other approaches such as
Bethe Ansatz and its algebraic formulation in
Quantum Inverse Scattering Method (QISM).

Whatever the original motivations and methods
were, it is  fair to say that the most
significant results in the theory
were obtained by bootstrap methods.
Exact S-matrices \cite{zz,kar}
and exact form factors \cite{kw,book} were found by this method.
Since the bootstrap calculation of the S-matrix
is based on semi-classical results from quantization
of solitons, and since the S-matrix defines the form factors through the
set of form factor axioms \cite{book}, one may imagine
that there is a direct path from
quantization of solitons to form factors.
Our attraction by this way of thinking does not mean,
of course, that we reject the achievements of QISM
which revealed the mathematical structure hidden
behind the integrable models and led to discovery
of quantum groups.

In recent years two important pieces of information were
added to the theory of sine-Gordon. The first of them is
the relation with the
perturbations of the
minimal models of conformal field theory (CFT)\cite{bpz,zam}.
The second is the discovery of the algebra of non-local charges, which
 is isomorphic
to the $q$-deformation of the universal enveloping algebra of the loop
algebra $\widehat{sl_2}$ \cite{bl2}.
Actually these two features are closely related:
the restricted sine-Gordon (RSG) model
coincides with the $\Phi _{1,3}$-perturbation of minimal CFT
 \cite{sm,bl1},
but the restriction is intimately connected with the
existence of the non-local symmetry.

In the present paper we shall show that the
results obtained by the bootstrap methods for sine-Gordon can be
understood directly by quantizing solitons.
Namely, we shall interpret the form factors of the restricted sine-Gordon
model as matrix elements in a quantum mechanical
$n$-soliton system
\footnote{A side motivation for this study was to learn how to
directly quantize solitons with potential applications to
theories where soliton-like solutions (monopoles) are known.}.
This will allow us
to underline the connection between profound
structures in the classical and quantum theories:
$\tau$-functions and separation of variables on the
one hand and the space of the local fields on the other
hand. Although we present the general structure for generic values
of the coupling constant, we will reconstruct the
sine-Gordon form factors only at the reflectionless points.
 We hope to return to the general case in another publication.

For each $n$-soliton solution
we shall introduce pairs of conjugated variables $A_i$ and
$P_i$ ($i=1\cdots ,n$),
which in the quantum case  satisfy Weyl commutation relations.
Every local operator $\CO$ can be considered as acting in
this $A$-representation, and therefore can be identified with a
certain operator $\CO (A,P)$. The typical formula for the matrix
element of $\CO$ between two $n$-soliton states can be
presented as
\debut
\vev{B'|\CO |B}=
\int\limits \Psi(A,B')^{\dag}\CO (A,P) \Psi(A,B)~d\mu(A),
\label{premier}
\fin
where $\Psi(A,B)$ is the wave-function of the state of
$n$ solitons with momenta  $B_1,\cdots ,B_n$.
The measure $d\mu (A)$ will include a specific weight
admitting a natural interpretation in the $n$-soliton symplectic
geometry.  We shall give explicit expressions for $\CO (A,P)$ corresponding
to the Virasoro primary fields.

In formula (\ref{premier}), the variables $A$ will be complex.
Choosing the integration domain is a non-trivial issue.
It  corresponds to choosing a real subvariety,
which specifies the configuration space of the theory.
This configuration space has a natural interpretation in
the {\it restricted} sine-Gordon model where it can
be understood as an analytical continuation of the sine-Gordon
or KdV configuration space. The hermitian conjugation $\dag$ is not
the naive complex conjugation inherited from the sine-Gordon
dynamics, but a more subtle one adapted to this choice
of real subvariety.

The local integrals of motion will be rewritten in the $A$-representation.
They are difference operators which can be expressed in
terms of ``quantum
$\tau$-functions".
We will describe how to separate the variables in
the associated Schr\"odinger equations.
The fact that these Schr\"odinger equations are difference equations
provides just enough room for the existence of the non-local
charges commuting with the Hamiltonians.
Recall that these charges do not allow direct classical limit.
The importance of coexistence of two commuting Weyl subalgebras was
pointed out in \cite{fadd}.

\section{The classical sine-Gordon theory.}
\subsection{Sine-Gordon solitons.}

In this section we introduce a few useful notations for the sine-Gordon (SG)
equation and its solutions.
Let $x_{\pm}=x\pm t$ be the light cone coordinates and
$\partial_{\pm} ={1\over 2}(\partial_x \pm\partial_t)$.
The sine-Gordon equation is:
\debut
\partial_+\partial_- \varphi = 2 \sin(2\varphi) \label{EAa}
\fin
It is convenient to introduce two $\tau$ -functions $\tau_\pm$,
in terms of which the SG equations can be rewritten in Hirota form.
The sine-Gordon field $\varphi$ is related to the $\tau $ -functions
by
$$ \exp(i\varphi)=\frac{\tau_-}{\tau_+} $$

Let us describe the $\tau$ -functions of the $n$-soliton solutions of
the SG equation.  Consider the function
\debut
\tau (X_1,\cdots ,X_n|B_1,\cdots ,B_n)= \det (1 + V  )
\label{deftau}
\fin
where $V$ is a $n\times n$ matrix with elements:
\debut
V_{ij}=2{B_i X_i\over {B_i+B_j}} \non
\fin
The $n$-soliton $\tau$-functions $\tau _{\pm}(x_-,x_+)$ are
written in terms of $\tau$ as follows
$$ \tau _{\pm}(x_-,x_+) = \tau _{\pm}(X(x_-,x_+)|B)$$
where
$$\tau _{\pm}(X|B) =\tau ({\pm}X|B)$$
The $x_{\pm}$-dependence of $X$ is quite simple:
\debut
X_i(x_+,x_-)=X_i\exp({2(B_ix_- +B_i^{-1}x_+)})
\label{evolution}
\fin
The quantities $X_i$ and $B_i$ are the parameters of the solitons:
$\b _i=\log (B_i)$ are the rapidities and $X_i$ are related to the positions.
For the sine-Gordon equation, they satisfy specific reality conditions.
For solitons or antisolitons, the rapidity $B$ is real and $X$ is
purely imaginary, i.e. $X=i\epsilon e^\gamma$  with $\epsilon=+1$ for a soliton
and $\epsilon=-1$ for an antisoliton.
We shall not consider ``breathers" in this paper but
for completeness it should be mentioned that they correspond to
pairs of complex conjugated rapidities $(B,\bar B)$
and positions $(X,-\bar X)$. Notice that these conditions are
preserved by the dynamics.

The sine-Gordon equation is a Hamiltonian system. The symplectic form
is the canonical one:
\debut
\Omega_{SG} =
\int_{-\infty}^{+\infty} dx\ \delta\pi(x) \wedge \delta\varphi(x) \nonumber
\fin
with $\pi(x)$ the momentum conjugated to the field $\varphi(x)$.
Above, $\delta$ denotes the variation on the phase space.
The space of $n$-solitons can be viewed as a $2n$-dimensional manifold embedded
into the infinite-dimensional phase space. The restriction of
the symplectic form on this finite-dimensional submanifold
gives the $n$-soliton symplectic form.
In the coordinates $X_i$ and $B_i$ it reads, cf eg.\cite{ft,bb1}:
\debut
\omega= \sum_{i=1}^n {d X_i \over X_i}\wedge {d B_i \over B_i}
+ \sum_{i<j} {4B_iB_j \over B_i^2-B_j^2 }
{d B_i \over B_i}\wedge {d B_j \over B_j}
\label{EAd}
\fin
The commuting conserved quantities are precisely the $B$'s.
In the following we shall always assume that the $B$ have
been ordered: $0<B_1<\cdots<B_n$.
A complete set of
commuting Hamiltonians $H_k$ can be chosen as
the set of elementary symmetric functions
\debut H_k = \sigma_k(B)\label{hm}\fin
We recall that the symmetric functions $\sigma_k(B)$ are defined by:
$\prod_j(z+B_j)=\sum_k z^{n-k}\sigma_k(B)$.
The local integrals of motion which are given on
$n$-soliton solutions by
$$I_{k}^{\pm}=s_{\mp (2k+1)}(B)\equiv \sum_{j=1}^n B_j ^{\mp (2k+1)}$$
can be expressed in terms of $H_i$. In particular
for the light cone components of the energy-momentum we
have $I_-\equiv I_1^{-}=H_1$, and $I_+\equiv I_1^{+}=H_n^{-1}H_{n-1}$.

The variable $Y_j$ canonically conjugated to $B_j$ is defined by
$Y_j = X_j~\prod_{k \neq j}\left({B_j -B_k \over B_j + B_k }\right)$.
The symplectic form is then written as
$\omega = \sum_j {d Y_j \over Y_j } \wedge {d B_j \over B_j}$.
The equations of motion are very simple
in the variables $\{Y,B\}$, which furthermore have the nice property of being
separated. Other set of variables have also been
introduced, in particular those which lead to Ruijsenaars
models \cite{rui,bb1}.
However, for the purpose of comparison with the existing
exact form factor formulae we need to introduce in the next section
still another set of variables.

A simple explicit expression for the $\tau$ -functions
can be obtained by expanding the determinant:
\debut
\tau (X|B) &=& 1 + \sum_{p=1}^n
\sum_{{ I \subset \{1,\cdots,n\} \atop |I|=p}}
\prod_{i<j \in I} \beta^2_{ij}(B)\cdot \prod_{i\in I} X_i
\label{tau1}\\
&=& 1 + \sum_j X_j + \sum_{i<j} \beta_{ij}^2(B) X_i X_j +
\cdots
\non
\fin
with $\beta_{ij}(B) = {B_i -B_j \over B_i + B_j}$.
In appendix B, we gather a few useful formulae
concerning these $\tau$-functions. In particular a
very useful formula is the recursion relation
satisfied by the $n$-solitons $\tau$-functions:
\debut
\tau ^{(n)}(X|B)= \tau ^{(n-1)}(X|B) +
\tau^{(n-1)}(\beta_{kn}^2(B)X_k|B)~X_n.
\label{recur1}
\fin

\subsection{The analytical variables.}

We now give a parametrization of the
$n$-soliton phase space in terms of new variables $\{A,B\}$.
The variables $B_j$ can be considered as poles of the Jost function
for the auxiliary linear problem, and the variables $A_j$
are zeroes of the Jost function. The importance of these
variables is better understood in the more general situation of
quasi-periodic finite-zone solutions from which
the soliton solutions are obtained by a limiting procedure.
In the finite-zone case the analogues of $B_j$ describe the
moduli of the hyper-elliptic spectral curve, while $A_j$ give the
divisor of zeroes of the Baker-Akhiezer function.
The general rule that the zeroes
of the Baker-Akhiezer function give the correct set of variables for
quantization was called the ``magic prescription" in \cite{skl}.
We discuss the relation to the finite-zone solutions in Appendix A.
Because of the nice algebro-geometrical meaning of these new variables
we call them the analytical variables.

The set of analytical variables $\{A,B\}$
is related to the variables $\{X,B\}$ introduced above by
\debut
X_j \cdot \prod_{k \neq j} \left({ B_j - B_k \over B_j + B_k }\right)
=  \prod_{k=1}^n \left({ B_j -  A_k \over B_j + A_k } \right),
\qquad for \qquad j=1,\cdots,n.
\label{XtoA}
\fin
This relation can be considered
as a system of equations for the symmetric functions
$\sigma_k(A)$ as functions of the $\{X,B\}$ variables.
The solution to this system is given in eq.(\ref{solA})
in Appendix B.

In these analytical variables the symplectic form becomes
\debut
\omega = 2 \sum_{k,j} {d A_j \wedge d B_k \over A_j^2 - B_k^2 }
\label{omAB}
\fin
We shall need the Liouville measure $\omega^n =\det (\om)dA_1 \wedge \cdots
\wedge dA_n \wedge dB_1 \wedge \cdots
\wedge dB_n$ with
\debut
\det (\om)\equiv \det\left({\inv{A_j^2-B_k^2}}\right)=
\frac{\prod_{j<k}(A_j^2-A_k^2)\prod_{j<k}(B_j^2-B_k^2)}
{\prod_{k,j}(A^2_j-B_k^2)} \label{sf}
\fin
It is useful to write the non vanishing Poisson brackets
\debut
\{ A_i, B_j \} = { \prod_{k \neq i} (B_j^2 - A_k^2)
\prod_{k \neq j} (A_i^2 - B_k^2 ) \over
\prod_{k \neq i} (A_i^2 - A_k^2) \prod_{k \neq j} (B_j^2 - B_k^2 )}
(A_i^2-B_j^2)
\non
\fin
Remark that the products in the right hand side can be written
in terms of cross-ratios. This is the first manifestation of the conformal
properties of these variables that will reappear in the following.

One can express the $\tau$-functions in terms of the variables $A_j$
and $B_j$. The result is the following surprisingly compact formula
\debut
\tau_+ &=& 2^n \left( \prod_{j=1}^n B_j\right)
 { \prod_{i < j} (A_i +A_j) \prod_{i <j} (B_i +B_j)
\over \prod_{i,j} (B_i + A_j) }
\label{tauAB1} \\
\tau_- &=& 2^n \left( \prod_{j=1}^n A_j\right)
 { \prod_{i < j} (A_i +A_j) \prod_{i <j} (B_i +B_j)
\over \prod_{i,j} (B_i + A_j) }
\nonumber
\fin
The symplectic form as well as the $\tau$-functions enjoy an
intriguing $A \leftrightarrow B$ duality.
The proof of these formulae is given in the Appendix B. They lead
to a formula expressing the sine-Gordon field in the $\{A,B\}$
variables:
\debut
e^{i \varphi } = {\tau_- \over \tau_+}
=  \prod_{j=1}^n \({A_j \over B_j}\)
\label{ephi}
\fin

Let us now introduce the variables $P_j$ conjugated to $A_j$
\debut
P_{j} = \prod_{k=1}^n\( {B_k-A_j  \over B_k+A_j }\),
\qquad for\qquad j=1,\cdots, n. \label{defP}
\fin
In terms of $A_j$ and $P_j$ the symplectic form takes the canonical form:
$$\omega =2\sum_{j=1}^n{dP_j\over P_j}\wedge {dA_j\over A_j} $$
We can express the
hamiltonians $H_k=\sigma_k(B)$ in terms of the variables $\{A,P\}$
using eqs.(\ref{defP}) as a linear system for the $\sigma_k(B)$.
Surprisingly, the solution of these equations can be written in
terms of the $\tau$-function as follows:
\debut
H_k= \sigma_k(B)={\tau _k(Z|A)\over \tau (Z|A)}
\qquad where \qquad
Z_j = (-1)^n P_{j} \prod_{k\neq j} \({A_j + A_k \over A_j - A_k } \)
\label{hkclass}
\fin
The functions $\tau_k$ are defined by
\debut
&\tau _k(Z_1,\cdots ,Z_n|A_1,\cdots ,A_n)=  \label{tauk} \\&=
\sum\limits _{i_1<i_2<\cdots <i_k} A_{i_1}A_{i_2}\cdots A_{i_k}
\tau(Z_1,\cdots ,-Z_{i_1}, \cdots,-Z_{i_k} ,\cdots ,Z_n|A_1,\cdots ,A_n)
\non
\fin
In particular $\tau (Z|A)=\tau _0(Z|A)$ and $\tau _-(Z|A)=
\bigl(\prod A_j\bigr)^{-1}\tau _n(Z|A) $.
We delay the proof of eq.(\ref{hkclass}) as it
turns out to be a limiting case of a more general quantum
formula which we shall prove in Appendix E.
It is convenient to introduce the generating function
of the $\tau_k$. Let $T(u)=\sum_{k=0}^n\, u^k \tau_k$.
It can again be expressed in terms of $\tau$-functions as:
\debut
T(u) =  \tau(Z(u)|A)~  \prod_{j=1}^n(1+uA_j)
\qquad with \qquad
Z_j(u) = \frac{1-uA_j}{1+uA_j}\, Z_j \label{tuclass}
\fin
This follows from the quantum relation eq.(\ref{genequant}) proved
in Appendix E.

Notice the unusual feature of our approach: the local
integrals of motion (\ref{hkclass}) are given in terms of
$\tau$-functions.

The equations of motion written in  the
variables $A_j$ are as follows:
\debut
\partial _- A_i=\{I_-,A_i\}=\prod\limits _j(A_i^2-B_j^2)
\prod\limits _{j\ne i} {1\over A_i^2-A_j^2} \non \\
\partial _+ A_i=\{I_+,A_i\}=\prod\limits _j {A_i^2-B_j^2\over B_j^{2} }
\prod\limits _{j\ne i}{A_j^2\over A_i^2-A_j^2}
\label{eqa}
\fin
These equations provide a particular case of
the general equations of motion for the divisor
of the zeroes of the Baker-Akhiezer function (see Appendix A).
This kind of
equations is commonly used in the theory of quasi-periodic solutions
of integrable equations; they go back to Neumann and Kowalevskaya.
One can show directly that (\ref{eqa}) together with (\ref{ephi})
imply the sine-Gordon equation \cite{har,paunov}.

To finish the discussion of the variables $A_i$ we have to
explain their trajectories.
For one soliton the condition $| e^{i\varphi} |=1$ shows that the
$A$-trajectory lies on the circle of radius $B$.
Under the classical SG dynamics of $n$-soliton solutions
every variable $A_k$ runs around a curve going in the
lower half-plane from $-B_k$ to $B_k$.
When $B_1\ll B_2 \ll \cdots \ll B_n$ the trajectories are
semi-circles, for finite $B_k$ they are getting deformed, but not too much.
For antisolitons these trajectories are replaced by their complex
conjugate.

\subsection{Reduced action and the relation to the KdV equation}

Our goal is to quantize of the (R)SG theory in the
soliton variables. The first step could be a semi-classical
quantization. To perform it for $n$-soliton
solutions, we need to compute the reduced action $\int ^q pdq$ since we are
restricting the system to the level of Hamiltonians, as in
the Maupertuis principle.  The symplectic form (\ref{sf})  can be rewritten as
$$\omega= d\alpha = \sum_{k=1}^n d\left(\log  \prod_j { B_j-A_k \over B_j+A_k }
\right) \wedge {d A_k \over A_k } $$
We discuss the relation of this symplectic form to the
general theory of analytical Poisson structures \cite{fm,nv}
in the Appendix A.
The  1-form $\alpha$ is defined up to an exact form $dF(A,B)$.
In this paper, we shall only consider the possibility of a function
$F(A,B)$ independent of $B$ (in order that $\alpha$ expends only on
$dA_j$, the coordinates) and of the special form $F(A_1,\cdots,A_n)=
\sum_k F_k(A_k) $ (to preserve the separability property of the $A_j$).
Hence the most general form of $\alpha$ is
\debut
\alpha = \sum_{k=1}^n \log \left( \prod_j { B_j-A_k \over B_j+A_k }
\right) \cdot {d A_k \over A_k } + \sum _k dF_k(A_k)
\label{1f}
\fin
The functions $F_k$ have to be fixed by additional considerations.

Now we face a serious challenge:
with these choices, the form $\alpha$ can not
arise in the full SG theory for the simple reason that
the reduced action $S(A_1,\cdots, A_n)$ constructed from $\alpha$
\debut
S(A_1,\cdots, A_n)=\sum_{k=1}^n
\int^{A_k} \log \left( \prod_j { B_j-A \over B_j+A }
\right) \cdot {d A \over A }  +F_k(A_k)
\nonumber
\fin
can not be made real along the sine-Gordon soliton $A_j$-trajectories
by any choice of $F_k$.
This circumstance looks very discouraging and the variables $A_j$
seem to be useless. But, as we shall later see, these are exactly
the variables in which the comparison with the quantum form factor
formulae is straightforward. We shall argue that the choices made for $\alpha$
correspond in fact to the restricted sine-Gordon (RSG) theory.
For this, we first need to describe the connection with the KdV equation.

Let us discuss briefly the relation between SG and KdV solitons.
The KdV equation allows soliton solutions in the form
$$  u(x_-)= \partial _- ^2\log\tau (x_-,0)$$
where $\tau$ is exactly the same as before.
In other words, the KdV soliton $\tau$-functions are
identical to the SG soliton $\tau$-functions but with all
the chiral coordinates $x_+$ set to zero.
The difference between the two cases lies in the reality conditions:
the variables $X_j$ which were imaginary for sine-Gordon become
real in the KdV case. Before going into the details of these
reality conditions, let us discuss the relation between the two cases
in the fully complexified situation.

In the variables $A_j$ the equations for KdV solitons are
exactly the same as for one chirality of SG (see Appendix A):
\debut
\partial _- A_i=\prod\limits _j(A_i^2-B_j^2)
\prod\limits _{j\ne i}{1\over A_i^2-A_j^2}
\label{eqa1}
\fin
The field $u$ reads in these variables as
\debut
u=2\left(\sum_j A_j^{2} -\sum_j B_j^{2}\right).
\label{ua}
\fin
It should be stressed that all the KdV fields are expressed
in terms of even powers of $A_j$.
The relation between the SG field $\varphi$ and the
KdV field $u$ is given by Miura transformation:
\debut
u(x_-)=-(\partial _-\varphi (x_-,0))^2-i\partial _-^2\varphi (x_-,0)
\label{mt}
\fin
It is a nice exercise to check directly that (\ref{ua}) follows
from $e^{i\varphi}=\prod_j { A_j\over B_j}$
using the equations (\ref{eqa1}).

As it has been said we are rather
interested in the RSG than in the SG theory. The RSG model
describe the $\Phi _{1,3}$-perturbation of the
minimal model of conformal field theory (CFT).
The reason why KdV equation appears in the context of
two-dimensional CFT is well known: the second
Poisson structure of KdV is a classical limit of the
operator product expansion for the light-cone component
of the energy-momentum tensor in CFT.
(For the minimal models of CFT the classical limit is
understood as the limit $c\to -\infty$.)
The second Poisson structure for KdV is:
\debut
\{u(x_-),u(x_-') \}=\delta '(x_- -x_-')(u(x_-) +u(x_-'))
+\delta '''(x_- -x_-')\label{vir}\fin
The KdV field $u$ is identified with the classical limit of $T_{--}$.
As explained in Appendix A, it turns out that the second
KdV Poisson structure restricted to the soliton manifold in the $\{A,B\}$
variables is
{\it identical} to the symplectic structure (\ref{omAB}) that
we have derived from the sine-Gordon theory. It is remarkable that the
conformal KdV Poisson structure appears in the (massive) sine-Gordon model
 when restricted to the soliton solutions. Again, this is only true
on the soliton submanifold.

Thus we are considering the light-cone hamiltonian picture, and
the lines $x_+=const$ are the space directions. The coordinate
$x_+$ has to be considered as time. Globally we can not
introduce the $x_+$ dynamics in KdV theory, but this
can perfectly be done on the $n$-soliton solutions:
the corresponding Hamiltonian is $I^+=\sum B_j^{-1}$ and the equations
of motion have the familiar form

\debut
\partial _+ A_i=\prod\limits _j{A_i^2-B_j^2\over B_j^{2} }
\prod\limits _{j\ne i}{A_j^2\over A_i^2-A_j^2}
\label{eqa2}
\fin

For real solutions the $x_+$-dynamics of $A_i$ is
organized as follows (see Appendix A). Each $A_i$  ($i\le n-1$) moves inside either of two
segments $(B_i,B_{i+1})$ and $(-B_{i+1},-B_i)$, the points $B_j$ and
$-B_j$ being identified (recall that the observables depend on $A_i^2$
only). The point $A_n$ moves in two segments $(B_n,\infty)$ and
$(-\infty ,-B_n)$ the points $\pm \infty$ being identified,
notice that $\infty$ is a regular point for the equations
(\ref{eqa2}).

Let us now normalize the 1-form (\ref{1f})
in order that it is real over $x_+$-trajectories. To do that
we have first to fix the branch of logarithms.
For further convenience we put $n$ cuts
over semi-circles in the upper-half plane connecting the points $B_j$
and $-B_j$ and require that the logarithm
$$\log \left( \prod_j { B_j-A \over B_j+A }
\right) $$
is real when $-B_1<A<B_1$. With these conventions
one easily figures out that the following choice of $F_k$
corresponds to a 1-form $\alpha$, real over $x_+$-trajectories:
\debut
\alpha = \sum_{k=1}^n \left(\log \left( \prod_j { B_j-A_k \over B_j+A_k }
\right) -k\pi i\right)\cdot {d A_k \over A_k }
\label{1fm}
\fin
The reduced action is real as well because the trajectories are real.

Returning to the relation with the sine-Gordon model, it is obvious
from (\ref{mt}) that real SG solutions do not correspond to
real KdV solutions.
One important consequence of the comparison with the
form factor formula is that we are actually considering
an analytical continuation of the sine-Gordon and KdV
dynamics. More precisely, the semi-classical quantization
exactly brings out the 1-form $\al$ as defined above in eq.(\ref{1fm}).
But the trajectories for $A_k$ lies on semi-circle of
radius $B_k<|A_k|<B_{k+1}$.
In other words, the 1-form needed for the quantization
of RSG model is obtained by analytical continuation of the
1-form  (\ref{1fm}). In this analytical continuation
we identify the variable $A_i$ which runs from $B_i$ to
$-B_i$ in SG case with the variable running inside
$(B_i,B_{i+1})\cup (-B_i,-B_{i+1})$ in KdV case
(we put $B_{n+1}\equiv\infty $).
Selecting  the trajectories in the complex $A$-plane is just
choosing the real phase space of the theory.

Following our logic
the form (\ref{1fm}) must be understood as restriction to $n$-soliton
submanifold of a globally defined 1-form related to the Poisson
structure (\ref{vir}). We are not able to define such a global
1-form, but we conjecture
that it exists. We can only refer to certain self-consistency checks
(such as reality of (\ref{1fm}) ) to support this conjecture.

\section{The SG form factors in absence of reflection of solitons.}

\subsection{ Sine-Gordon versus Restricted Sine-Gordon.}

The sine-Gordon equation follows from the action:
$$ S={\pi\over\gamma}\int \CL \ d^2 x,\qquad
\CL=(\partial _{\mu}\varphi)^2+
m^2 (\cos (2\varphi)-1)  $$
where $\gamma$ is the coupling constant, $0<\gamma<\pi$.
The free fermion point is at $\ga=\frac{\pi}{2}$.
In the quantum theory, the relevant coupling constant is:
$$\xi ={\pi \gamma\over \pi -\gamma}. $$
We shall always use the constant $\xi$, which plays the role
of the Planck constant.
Only the mass is renormalized but not the coupling
constant $\ga$ \cite{col}.

The SG theory is invariant under the quantum affine loop algebra
$U_{\hat q}(\hat {sl_2})$ with $\hat q = \exp(i\frac{2\pi^2}{\xi})$.
This symmetry algebra is generated by the topological charge
and four non-local charges of spin $s=\pm\frac{\pi}{\xi}$.
(By convention, $\hat q$ here is the inverse square of
that of ref.\cite{bl2}.)

The canonical stress tensor of the SG theory is
$T_{\mu\nu}^{SG}= (\d_\mu\varphi)(\d_\nu\varphi)
-\frac{g_{\mu\nu}}{2} \CL $. The SG theory
contains two subalgebras of local operators
which, as operator algebras are generated by $\exp (i\varphi)$
and $\exp (-i\varphi)$ respectively. Let us concentrate on one
of them, say the one generated by $\exp (i\varphi)$. It is known that
this subalgebra can be considered independently of the rest of the
operators as the operator algebra of the theory with the
modified energy-momentum tensor:
\debut
T^{mod}_{\mu \nu}=T^{SG}_{\mu \nu}+i\pi \sqrt {6\over \xi(\pi +\xi)}
\epsilon _{\mu,\mu '}\epsilon _{\nu,\nu '}
\partial _{\mu '}\partial _{\nu '}\varphi  \label{tmod}
\fin
This modification changes the trace of the stress tensor which is now:
$$T^{mod}_{\mu \mu}= m^2  \exp (2i\varphi) $$
This modification corresponds to the restricted Sine-Gordon theory (RSG).
For rational ${\xi\over\pi}$ it describes
the $\Phi _{1,3}$-perturbations of the minimal models of CFT,
but it can be considered for generic values of $\xi$ as well.
The central charge and the dimension of the operator $\Phi _{1,3}$
are given by
$$c=1-{6\pi ^2\over\xi(\pi +\xi)},\qquad
\Delta _{1,3}={\xi-\pi\over\xi+\pi}$$

Modifying the stress tensor as in eq.(\ref{tmod}) modifies the Lorentz
boost and hence the spin of the non-local charges. Under this modification
two of these charges become spinless. Together with the
topological charge they then form a representation of the finite
quantum algebra $U_{\hat q}({sl_2})$. The restriction of the
Sine-Gordon model consists in gauging away this symmetry subalgebra.
The physical states of the RSG model are annihilated by
these spin-less non-local charges.
The physical operators of the RSG model
are those which commute with these charges. In particular,
$e^{i\varphi}$ which commute with them is a physical
operator, but $e^{-i\varphi}$ is not since it does not
commute with these charges.

The asymptotic states of the RSG theory are constructed as follows.
Consider the states in the SG theory containing $n$ solitons and
$n$ anti-solitons:
$$ |\b _1,\b _2,\cdots ,\b _{2n}\rangle _{\epsilon _{1}, \epsilon _{2}
\cdots ,\epsilon _{2n}}$$
where $\b _j$ are the rapidities of particles, and $\epsilon _{j} $ is
$+$ or $-$ for soliton or anti-soliton respectively. For the  RSG theory
one introduces the states
\debut
 |\b _1,\b _2,\cdots ,\b _{2n}\rangle\rangle _{\epsilon _{1}, \epsilon _{2}
\cdots ,\epsilon _{2n}}
=\exp ({\pi\over 2\xi}\sum_j \epsilon _{j} \b _j)
|\b _1,\b _2,\cdots ,\b _{2n}\rangle _{\epsilon _{1}, \epsilon _{2}
\cdots ,\epsilon _{2n}}\fin
The extra factor in $|\b _1,\b _2,\cdots ,\b _{2n}\rangle\rangle$
is an echo of the modification (\ref{tmod}) of the stress tensor.
In the new basis the S-matrix becomes manifestly invariant
under $U_{\hat{q}}(sl_2)$.
The asymptotic states of the RSG model are then the
$U_{\hat{q}}(sl_2)$-scalar in the Hilbert space span by the
states $|\b _1,\b _2,\cdots ,\b _{2n}\rangle\rangle$.

In the restricted theories one generally can not introduce a
positively defined Hermitian structure.
Obviously the SG hermitian conjugation $*$ maps the RSG model
into the symmetric restricted model
constructed from $\exp (-i\varphi)$. On the other hand the
SG charge conjugation ($\varphi\to -\varphi$) is also broken in the RSG model.
However, one can introduce an anti-linear involution for the RSG model
as the combined CT reflection. For any local operator the combined
transformation (which we denote by $\dag$) corresponds to
hermitian conjugation and reflection $\varphi\to -\varphi$.
This operation $\dag$ does not give a positively defined scalar
product for the SG theory, but it does not lead to contradiction
if one stays within the RSG theory. In particular, since
the modified stress tensor $T^{mod}_{\mu\nu}$ is
hermitic in the RSG theory,  one has:
\debut
\left(e^{i\varphi}\right)^{\dag}=e^{i\varphi} \label{basicss}
\fin
This is a simple but fundamental remark.

\subsection{Form factor formulae.}

In what follows we shall consider the case $\xi ={\pi\over \nu}$ for
$\nu =1, 2,\cdots$, when the reflection of solitons and
anti-solitons is absent. In these cases, the two parameters
$q$ and $\hat q$ are
\debut
q= e^{i\frac{\pi}{\nu}}\qquad,\qquad \hat q=1 \non
\fin
The S-matrix is diagonal and given by
$$S(\b)=\prod\limits _{j=1}^{\nu -1}{\sinh{1\over 2}(\b +{\pi i\over\nu}j)
\over\sinh{1\over 2}(\b -{\pi i\over\nu}j)}= \prod_{j=1}^{\nu -1}
\left( {B q^j -1 \over B -q^j } \right)$$
We shall use the following notations:
$$B=\exp (\b),\qquad b=\exp ( {2\pi \over\xi} \b)=\exp (2\nu \b)$$

Consider any local operator ${\cal O}(x)$ for RSG. Its form factors
are defined by
\debut
f_{\cal O}(\b _1,\b _2,\cdots ,\b _{2n})_{\epsilon _{1}, \epsilon _{2}
\cdots ,\epsilon _{2n}}=
\langle\langle 0|\CO (0)
|\b _1,\b _2,\cdots ,\b _{2n}\rangle\rangle _{\epsilon _{1}, \epsilon _{2}
\cdots ,\epsilon _{2n}}
\fin
The form factors are given by the formulae
\debut
f_\CO (\b _1,\b _2,\cdots ,\b _{2n})_{- \cdots -+\cdots +}&=&
e^{(-{1\over 2}(\nu(n-1)-n)\sum_j \beta_j)}
\prod\limits _{i<j}\zeta (\b _i-\b _j)
\prod\limits _{i=1}^n \prod\limits _{j=n+1} ^{2n}
{1\over\sinh{\pi\over\xi}(\b _j-\b _i -\pi i)}
\non \\
&&\hskip +3cm
\times \widehat{f}_\CO (\b _1,\b _2,\cdots ,\b _{2n})_{-\cdots -+\cdots +}
\label{ff}
\fin
The function $\zeta (\b)$ is regular in the
physical strip $0< Im ~ \beta ~< \pi$, and satisfies $\zeta (-\beta) =
 \zeta(\beta -2i\pi)=S(\beta) \zeta (\beta)$. It can be found in \cite{book} . We shall not
consider this prefactor in eq.(\ref{ff}) since it is the
same for all operators: it is related to normalization of
the wave function which we hope to discuss sometime.
The most interesting part of the form factor is given by 
\debut
\widehat{f}_\CO (\b _1,\b _2,\cdots ,\b _{2n})_{-\cdots -+\cdots +}&=&
\non\\
&&\hskip -5cm
={1\over (2\pi i)^n}\int\limits _C dA_1\cdots \int\limits _C dA_n
\prod\limits _{i=1}^n \prod\limits _{j=1} ^{2n} \psi (A_i,B_j)
\prod\limits _{i<j} (A_i^2-A_j^2)
\ L_\CO (A_1,\cdots ,A_n|B_1,\cdots ,B_{2n})
\prod\limits _{i=1}^n a_i^{-i} \label{ints}
\fin
Some comments are needed for this formula.
The function $\psi (A,B)$ is given by
\debut
 \psi (A,B)=\prod\limits _{j=1}^{\nu -1}(B -Aq^{-j} )
\label{defpsi}
\fin
This function satisfies the difference equation:
\debut \psi(Aq,B)=\left({B-A\over B+qA}\right)~ \psi(A,B)
\label{eqpsi}\fin
As usual we define $a=A^{2\nu}$.
The contour $C$ is drawn around the point $A=0$.
Notice that the right hand side of the formula (\ref{ints}) does not
depend on the partition of the particles into solitons
and anti-solitons, this is a peculiarity of reflectionless case.

Different local operators are defined by different
functions $L_\CO (A_1,\cdots ,A_n|B_1,\cdots ,B_{2n}) $.
These functions are symmetric polynomials of $A_1,\cdots ,A_n$.
For the primary operators $\Phi _{2k}=\exp (2k i\varphi )$
and their Virasoro descendants, $L_\CO$ are symmetric Laurent
polynomials of $ B_1,\cdots ,B_{2n} $. For the
primary operators $\Phi _{2k+1}=\exp ((2k+1) i\varphi)$,
they are symmetric Laurent polynomials of $ B_1,\cdots ,B_{2n} $
multiplied by $\prod B_j ^{1\over 2}$.
Our definition of the fields $\Phi _{m} $ is related to the notations coming
from CFT as follows: $\Phi _{m}$ corresponds to $ \Phi _{1,m+1}$.
The explicit form of the polynomials $L_\CO$ for the primary operators
is as follows
$$ L_{\Phi _m}(A_1,\cdots ,A_n|B_1,\cdots ,B_{2n})
=\prod\limits _{i=1}^n A_i^m
\prod\limits _{j=1}^{2n} B_j^{-{m\over 2}} $$
We explain in Appendix C that this definition agrees with
the formulae for the form factors of the operators $\Phi _1$ and
$\Phi _2$ given in \cite{book}.
We shall return to this definition in the following sections.

Usually the formulae for the form factors are written in
a slightly different way \cite{book,count}. First, the
integrals are $(n-1)$-fold, second, instead of the
polynomials $\prod\limits  (A_i^2-A_j^2)
L_\CO (A_1,\cdots ,A_n|B_1,\cdots ,B_{2n}) $ under the
integral, one usually has polynomials of the type
$\prod\limits  (A_i -A_j)
\widehat{L}_\CO (A_1,\cdots ,A_{n-1}|B_1,\cdots ,B_{2n}) $
with the limitation $deg_{A_i}(\widehat{L}_\CO) \le n$. The
conventional form of the integrands is important for
generalization to the case of arbitrary
coupling constant because, in this case, the integrals
become more complicated and the above limitation is needed for
convergence. In Appendix C we explain briefly
how to rewrite formula (\ref{ints}) in the conventional way.

There are two important facts that we learn
from the calculations of Appendix C.

1. The substitution into (\ref{ints}) of the polynomial
$L_I(A_1,\cdots ,A_n)=1$ which corresponds to the unit operator,
gives zero because there is no simple pole in the contour integral, in
agreement with the fact that
 the matrix element of the unit operator
between the vacuum and an excited state vanishes.
Also if we try to substitute into (\ref{ints})
the functions
$$ L_{\Phi _{-m}}(A_1,\cdots ,A_n|B_1,\cdots ,B_{2n})
=\prod_j A_j ^{-m}\prod_j B_j^{{m\over 2}} $$
which must correspond to the operators
$\Phi _{-m}=\exp (-m i\varphi ) $
the integral vanishes. This means exactly that
our formulae suit rather the RSG than the SG model.
This does not mean that the formulae for the form factors
of $\Phi _{-m}$ do not exist: they are obtained by SG
charge conjugation, we want to say only that these
formulae can not be obtained by putting $ L_{\Phi _{-m}}$
into the integral formula.

2. Consider any polynomial $M(A_1;A_2,\cdots ,A_n)$
anti-symmetric with respect to $A_2,\cdots ,A_n$.
The value of the integral (\ref{ints}) does not change if we add to
$\prod (A_i^2-A_j^2)L_\CO (A_1, \cdots ,A_n)$ an ``exact form''
\debut
\sum\limits _k (-1)^k
\left(
M(A_k;A_1,\cdots ,\widehat{A_k},\cdots ,A_n) \prod_j (B_j+A_k)-
q M(q A_k;A_1,\cdots ,\widehat{A_k},\cdots ,A_n) \prod_j (B_j-A_k)\right)
\label{zero}
\fin
where $\widehat{A_k}$ means that $A_k$ is omitted.

\section {The semi-classical analysis.}

\subsection{The semi-classical analytical quantization.}

In this section we show that the exact formulae
for the form factors in the semi-classical limit
can be obtained from the semi-classical quantization of solitons
in the analytical variables.
We shall proceed to the exact quantization of solitons in
the next section.

We shall consider the matrix elements of the operator $\CO$
 calculated between
two $n$-soliton states, instead of the ones
calculated between the vacuum and the state with $n$ solitons
and $n$ anti-solitons.
 This is because they have a more direct semi-classical
interpretation.
Crossing symmetry relates the
connected parts of these two kinds of matrix elements.
The essential piece
of the $n$ solitons to $n$ solitons form factor
is given by
\debut
\widehat{f}_\CO (\b _1',\cdots ,\b _{n}'|\b _1,\cdots ,\b _{n})&=&
{1\over (2\pi i)^n}\int\limits _C dA_1\cdots \int\limits _C dA_n
\prod\limits _{i=1}^n \prod\limits _{j=1} ^{n}
\psi (A_i,-B_j')\psi (A_i,B_j)
\prod\limits _{i<j} (A_i^2-A_j^2)
\prod\limits _{i=1}^n a_i^{-i}
\non\\
&&\hskip 2cm \times
L_\CO (A_1,\cdots ,A_n|-B_1',\cdots ,-B_{n}', B_1,\cdots ,B_{n}) \label{n+n}
\fin
where $\psi(A,B)$ is defined in eq.(\ref{defpsi}).
We would like to present this formula as
\debut
 \int\limits_{\cal D} \Psi(A,B')^{\dag}\CO (A,P) \Psi(A,B)W(A)dA
\label{repff}
\fin
where the  various terms should be interpreted as follows.
The function $\Psi(A,B)$ is the $n$-soliton wave function
in the Schr\"odinger $A$-representation. The scalar product
in the Hilbert space is written in terms of a non trivial weight $W(A)$
and a possibly non standard conjugation $\dag$. The integration domain ${\cal
D}$ has to be compared to the configuration space of the classical
theory.
The quantity $\CO (A,P)$ is the operator $\CO$
realized in terms of $A$ and the conjugated variables $P$.
The detailed discussion of the operators $A$ and $P$
will be given in the next section.
Let us first make a semi-classical step in this direction.

Consider the semi-classical approximation of the
function $\psi(A,B)$. Asymptotically we have (see \newline Appendix D)
\debut
\psi(A,B)\sim _{\nu\to\infty}{B^{\nu}\over \sqrt{B^2 -A^2}}
\exp\left({\nu\over i\pi}\int\limits _0^A\log\left(B-A\over B+A\right)
A^{-1}dA\right)
\label{semipsi}
\fin
where the integral is taken over a contour which does not
cross the cut going along the semi-circle in the upper-half plane
from $B$ to $-B$. The logarithm is
real when $-B<A<B$. An important property
of the reflectionless situation is that the integrals over
different contours of this kind
give the same result when substituted into the exponent.
Indeed we have,
$$ \int_\gamma \log\left(B-A\over B+A\right)
A^{-1}dA =-2\pi ^2 $$
where the contour $\gamma$ is drawn around the cut.
Therefore, when $\nu$ is integer there is no ambiguity in the
exponential in eq.(\ref{semipsi}).
This is similar to the topological ambiguity of the WZNW action.

Let us now construct the semi-classical wave-function in the
$A$-representation for given values of $B$.
The general rule is \cite{landau}:
\debut
\tilde{\Psi}(A)\simeq \left(d\mu (A)\right)^{1\over2}
\exp{{1\over i\hbar}S(A)}
\label{land}
\fin
where $d\mu (A)\equiv \omega^n$ is the Liouville measure at the point $A$.
Let us give some explanation. Integrable models
provide  examples of a situation which is usually described in textbooks
only for the case of one degree of freedom (i.e. when there is
only one integral of motion: the energy). In our case in order to
construct the semi-classical wave-function in the $A$-representation,
we have to consider the polarization corresponding to the $A$ variables
 and the integrals of motion $B$. In these
variables the Liouville measure can be written as
$$d\mu (A,B)=[\det~\om(A,B)]~dA_1\wedge\cdots\wedge dA_n\wedge
dB_1\wedge\cdots\wedge dB_n$$
 So $(d\mu (A))^{1\over 2} $
has to be understood as $(\det \om (A,B))^{1\over 2}$.
In this measure the terms which depend only on $B$ will be omitted
since they correspond to the
normalization of the wave-functions
which we do not consider.

In eq.(\ref{land}), $S(A)$ is the semi-classical action. We choose as
semi-classical action the analytical continuation of the KdV reduced action
discussed in the previous section, which
in the analytical variables $A$
is $S(A) = \sum_k s_k(A_k)$ with
\debut
s_k(A_k) = \int\limits ^{A_k}_0 \log \left( \prod_j { B_j-A \over B_j+A }
\right) \cdot {d A \over A } -\pi i k\log(A_k)
\label{act}
\fin
where the integrals are again taken in the plane with cuts
going along the semi-circle
from $B_j$ to $-B_j$ for all $j$. The ambiguity in the
definition of the contours is irrelevant for semi-classical quantization at
the reflectionless points
as it has been explained above.
For the case of generic coupling constant we can not
manage with the variables $A_i$, instead the variables
$\alpha _i=\log (A_i)$ will have to be considered. We shall return to
this point in future publication.

We now have all the necessary ingredients: the classical action and the
Liouville measure. So, up to normalization depending only on the $B_j$,
the semi classical wave function is
\debut
&\tilde{\Psi} (A_1,\cdots ,A_n|B_1,\cdots,B_n)\simeq
\label{bigpsi} \\
&\left(\prod\limits _{i<j}(A_i^2-A_j^2)\right)^{1\over2}
\left(\prod\limits _{i=1}^n\prod\limits _{j=1}^n
(A_i^2-B_j^2)\right)^{-{1\over2}}
\exp\left({\nu\over i\pi }\left(\sum\limits _{k=1}^n
\int\limits _0^{A_k} \log \left( \prod_j { B_j-A \over B_j+A }
\right) \cdot {d A \over A } -\pi i k\log (A_k)\right)\right) \non
\fin
The Planck constant is identified with $\xi={\pi\over\nu}$.
We shall divide $ \tilde{\Psi} $ into the wave function $\Psi$
and a $B_j$-independent piece which will be put into the
integration measure:
$$ \tilde{\Psi} (A_1,\cdots ,A_n|B_1,\cdots,B_n)=
(W(A_1,\cdots ,A_n))^{1\over 2}
 \Psi (A_1,\cdots ,A_n|B_1,\cdots,B_n) $$
where
\debut
W(A_1,\cdots ,A_n)~~~~~&=&
\prod\limits _{i<j}(A_i^2-A_j^2) \prod\limits _{k=1}^n a_k^{-k}
\non \\
\Psi(A_1,\cdots ,A_n|B_1,\cdots,B_n) &=& ~~
\prod_{i=1}^n \prod_{j=1}^n \psi(A_i,B_j)
\non
\fin
where we recall that $a_k=A_k^{2\nu}$.

Notice that choosing the reduced action as we did, i.e. as the analytical
continuation of the KdV action, was  crucial for producing the factors
$ \prod_k a_k^{-k}$ in the weight $W(A)$. Identifying $\Psi(A,B)^{\dag} =
\Psi(A,-B)$ and the integration domain ${\cal D}$ with the contour integrals
over $C$ we see that the form factors in the semi classical limit admit
a representation as in eq.(\ref{repff}).

It remains to find a more physical and geometrical
interpretation of the integration domain and of the
conjugation $\dag$ in the semi-classical regime.
Under the classical dynamics of $n$-soliton solutions,
every variable $A_k$ runs around a curve going in the
lower half-plane anti-clockwise
from $-B_k$ to $B_k$.
As we already know there are two difficulties: \newline
1. The action $s(A_k)$ is not real along the classical
trajectory. \newline
2. The trajectory itself is complex, and its
exact geometry depends on $B_k$ which must become
the spectral data after the quantization.\newline
On the other hand the exact quantum formula show that the
integration domains should be small circle around the origin.
This suggests the following prescription for fixing these
problems. \newline
1. Because of quantum-mechanically possible penetration through
the barrier one has to consider the closed trajectories, i.e.
to add to the classical trajectory of $A_k$ the piece running in
upper half-plane anti-clockwise from $B_k$ to $-B_k$.
Classically this extra piece corresponds to an anti-soliton trajectory.
\newline
2. In order to eliminate the geometrical dependence
of the trajectory of $A_k$ on $B_j$ (leaving the topological one)
we replace it by arbitrary closed curve along which $B_k<|A_k|<B_{k+1}$.
There is one more reason for considering the configuration space
as the one composed of closed cycles. If we regularize the theory
putting it onto large but finite interval with periodic boundary conditions
the trajectories of $A_j$ will become the closed cycles of the
same type as described above (see Appendix A).\newline
3. Last but not least, the reality condition is taken as follows:
\debut
s_k(A_k)=\overline{s_k(\overline{A_k}) }\equiv s(A_k)^{\dag}
\label{real}
\fin

Let us explain that this condition is satisfied by (\ref{act}).
If we take
$$\int\limits ^{A_k}_0 \log \left( \prod_j { B_j-A \over B_j+A }
\right) \cdot {d A \over A }
\qquad and
\qquad \int\limits ^{\overline{A_k}}_0 \log \left(
\prod_j { B_j-A \over B_j+A }
\right) \cdot {d A \over A }$$
over conjugated contours, then
(\ref{real}) would be true  if the integrands were
continuous over these contours. However, we defined the
logarithms in such a way that they have cuts in the
upper half plane, and to reach $A_k$ we have to cross
$k$ cuts. The addition of $-\pi i k\log(A_k)$
in (\ref{act}) is needed to compensate the corresponding
 discrepancy.

The function $\Psi ^{\dag}$ is constructed from $s^{\dag}$.
The choice of the conjugation $\dag$ ensures that
$\Psi(A|B')^{\dag}=\Psi(A|-B')$ as it should be.

Let us discuss the condition (\ref{real}). This
condition provides the usual reality condition
$$s_k(A_k)=\overline{s_k(A_k) } $$
if $A_k$ is real and such that $B_k<A_k<B_{k+1}$. The
latter segment is the trajectory of $A_k$ which
corresponds to real $n$-soliton solutions
of KdV equation. So, we are dealing with an analytical
continuation of KdV mechanics,
as it has been explained in the Section 2.3.
This fits perfectly with our assumption that we are actually
doing not SG but RSG model, which has to be considered as
a continuation of KdV. Also the fact that the
involution (\ref{real}) connects the soliton and the
anti-soliton part of the trajectory has a physical
explanation:
as we already said  RSG is neither C nor T invariant,
so one has to consider the combined CT invariance.
The involution (\ref{real}) corresponds exactly to the combined CT reflection.

It now is clear that the formula (\ref{n+n}) allows a semi-classical
interpretation by means of this semi-classical wave function.
We stress once again that this semi-classical analysis forces us
to introduce the analytical continuation (\ref{real}).
The insertion of a local operator corresponds to the
insertion of a polynomial $L_\CO$ which can also be understood
from the expression of the classical local observables in $n$-soliton
variables. We shall return to this point in the section on exact quantization.

Let us emphasize one circumstance.
In order to put together two
semi-classical wave-functions $\Psi(A,B)$ and
$\Psi(A,B')^{\dag}$ into the matrix element, one has to
be sure that the contours of $A_k$ can be identified.
Recall that for these two wave functions the contours are
inside $B_k<|A_k|<B_{k+1}$ and $B_k'<|A_k|<B_{k+1}'$ respectively.
This shows that the semi-classical quantization
works nicely if the soliton states are not very far from
each other in the following sense. The sets $B_j$ and
$B_j'$ are ordered $B_k<B_{k+1}$ and $B_k'<B_{k+1}'$.
The semi-classical formulae are applicable if in addition
$B_k'<B_{k+1}$ and $B_k<B_{k+1}'$. In the next section we
shall discuss this point further.

\subsection{Stationary point calculation of the semi-classical integrals.}

In this section we shall show that the semi-classical integrals
for the form factors are actually given by stationary
point contributions. This result, which is interesting by itself,
will also provide some justification of the compactification
of the configuration space made before.

It is convenient to return to the form factor taken
between the vacuum and the state with $n$ solitons
and $n$ anti-solitons. One can show that the semi-classical limit
allows the necessary analytical continuation.
Let us consider the exact quantum formula:
\debut
\widehat{f}_\CO (\b _1,\b _2,\cdots ,\b _{2n})_{-\cdots -+\cdots +}&=&
\non\\
&& \hskip -5cm
={1\over (2\pi i)^n}\int\limits _{C_1} dA_1\cdots \int\limits _{C_n} dA_n
\prod\limits _{i=1}^n \prod\limits _{j=1} ^{2n} \psi (A_i,B_j)
\prod\limits _{i<j} (A_i^2-A_j^2)
\ L_\CO (A_1,\cdots ,A_n|B_1,\cdots ,B_{2n})
\prod\limits _{i=1}^n a_i^{-i}\label{xxx}
\fin
In quantum formula the contours of integration are
arbitrary. However, to perform the classical limit
we need to take them as follows: if $A_k\in C_k$ then $B_{2k}<|A_k|<B_{2k+1}$.
The point is that
the asymptotic of $\psi (A,B)$ is not a single-valued function
of A, so one can not move the contours of integration
after the asymptotic of the integrand is calculated.
The above prescription for the contours agrees with the semi-classical
considerations of the previous section.
The justification of this choice of contours follows from further
stationary phase calculations.
The momentum $B_i$ corresponds
to either soliton or anti-soliton. The matrix element is
related to that considered in the previous section by analytical
continuation. The concluding remarks of the previous section
show that it is exactly the choice of cycles described above
which corresponds to the semi-classical matrix element, and
the momenta $B_j$ are partitioned into pairs $B_{2i-1}, B_{2i}$
such that in every pair we find one soliton and one anti-soliton.

The asymptotic of the integrand is written down
using the formulae from the previous section.
Let us consider the integral with respect to $A_k$.
It contains the divergent exponent:
$$ \exp\left({\nu\over i\pi}\left(\int\limits _0^{A_k} \log\prod\limits _j
\left(B_j-A_k\over B_j+A_k\right)dA_k-2\pi ik
\log (A_k)\right)\right) $$
where the last term comes from $\prod_k a_k^{-k} $.
One has to consider the stationary phase point at $A=-iF_k$
which solves the equation
$$ \log\prod\limits _j\left(B_j-iF_k\over B_j+iF_k\right)=2\pi ik  $$
It is easy to see that all $F_k$ are real and positive, moreover
$B_{2k}<F_k<B_{2k+1}$ for $k=1,\cdots n-1$, $F_n=\infty$.
The calculation of the second derivatives shows that the
integrals over $A_k$ for $k=1,\cdots n-1$ are given by
stationary point contributions coming from $-iF_k$ if
the contour of integration goes trough the point $F_k$
being topologically equivalent to a circle
lying inside the domain $B_{2k}<|A_k|<B_{2k+1}$.
Obviously the contours $C_k$ from (\ref{xxx}) can be
drawn like this.
The point $\infty$ for the $n$-th integral is
not a true stationary point, but for  the
$n$-th integral the divergent exponent vanishes when
$A_n\to\infty$, and the integral is sitting on the residue
just like in the exact quantum calculation (Appendix C).

The equations for the stationary points can be
summarized into the Bethe Ansatz type equation
$$ \prod\limits _j\left(B_j-iF\over B_j+iF\right)=1  $$

Consider any polynomial $M(A_1;A_2,\cdots ,A_n)$
anti-symmetric with respect to $A_2,\cdots ,A_n$.
The main contribution
into the asymptotic of the integral (\ref{xxx}) does not change if we add to
$\prod (A_i^2-A_j^2)L_\CO (A_1, \cdots ,A_n)$ the expression
\debut
\sum\limits _k (-1)^k
M(A_k;A_1,\cdots ,\widehat{A_k},\cdots ,A_n) (\prod (B_j+A_k)-
 \prod (B_j-A_k))
\label{zerocl}
\fin
because it vanishes
on the stationary point if $k\le n-1$, and cancels residue if
$k=n$. Comparing this formula with exact quantum formula (\ref{zero})
we observe the following nice circumstance. The semi-classics is
of course not exact like it happens in geometrical quantization,
however one can think of the exact quantum relation (\ref{zero}) as
of deformation of (\ref{zerocl}) which in turn is nothing
but the consequence of the equation for the stationary points.
The factorization over (\ref{zerocl}) and (\ref{zero}) leave the same number
of independent polynomials of $A_k$.

The stationary phase calculation fits nicely with the classical
picture. The point $-iF_k$ lies in the region acceptable by
analytical continuation from the
classical soliton trajectory. This fact gives an important justification of
our semi-classical methods.

\section{Exact analytical quantization of solitons.}

\subsection{Hilbert Space and Hermitian structure.}

In this section we do not specify $\xi$ to the
values ${\pi\over\nu}$ until the last subsection
in which
we shall reconstruct the RSG form factors.
Consider the variables $\al _j=\log (A_j)$ and
$\pi _j =\log (P_j)$. The symplectic form is canonical
in terms of these variables:
$$\omega =2\sum_{j=1}^n d\pi_j\wedge d\al _j $$
We quantize them in a canonical way:  $\al_j$ act
by multiplication and $\pi_j$ by differentiation, i.e.
$\al _j\to\al _j$ and $\pi _j\to i\xi{\partial\over
\partial \alpha _j}$.
Here $\xi$ plays the role of Planck constant.
The operators $A$ and $P$ are defined as:
$$A_j=\exp (\alpha _j),\qquad P_j=\exp (i\xi{\partial\over
\partial \alpha _j} )$$
They satisfy Weyl commutation relations with $q=\exp(i\xi)$:
\debut
P_j\cdot A_j &=& q\; A_j\cdot  P_j,  ~~~~~~~~~~~~
with \qquad q= e^{i\xi} \label{comrelAP}\\
P_k\cdot A_j &=&  A_j\cdot  P_k, ~~~~~
\qquad j\not= k\non
\fin
This definition is rather formal since we have not yet defined
the Hilbert space of functions of $\alpha _j $. Staying on the
same formal level one realizes the following important circumstance
\cite{fadd}. There is another pair of operators
$$a_j=\exp ({2\pi\over\xi}\alpha _j),\qquad
p_j=\exp (2\pi i{\partial\over
\partial \alpha _j} ) $$
which represent the Weyl algebra but with the dual quantum parameter
$\hat{q}=\exp({2\pi ^2i\over\xi})$, associated to the
non-local symmetry algebra $U_{\hat q}(\hat {sl_2})$:
\debut
p_j\cdot a_j &=& \hat{q}^2\; a_j\cdot  p_j, ~~~~
with\qquad \hat{q}=\exp({2\pi ^2i\over\xi})\non\\
p_k\cdot a_j &=&  a_j\cdot  p_k, ~~~~
\qquad j \not= k \label{comdual}
\fin
The operators $a_j$ and $p_j$ commute with $A_j$ and $P_j$.
The existence of two dual algebras will be crucial in the following.

All these operators act on wave functions $\Psi(\al)$.
In the reflectionless case $\Psi(\al)$
is a single-valued function of $A$'s so we shall write
$\Psi(A)$ in that case.
The operators $A_j$ act by multiplication whereas $P_j$ act by
shifting the argument of $\Psi$ by $i\xi$:
\debut
P_j \Psi(\al _1,\cdots,\al _j ,\cdots,\al _n)
= \Psi(\al _1,\cdots,\al _j +i\xi ,\cdots,\al _n)\non
\fin
To complete the representation of the canonical commutation relations
in the Hilbert space, we also need to introduce the scalar product.
In order to take into account the reality condition
$(e^{i\varphi})^{\dag}= e^{i\varphi}$, specific to the RSG model,
we define the scalar product in rather unusual way.
Let $\Psi_1(A)$ and $\Psi_2(A)$ be two wave functions, then:
\debut
\vev{\Psi_1|\Psi_2}= \int_C d\al _1\cdots\int_C d\al _n \prod_i A_i
\prod_{i<j}(A^2_i-A^2_j)~ \Psi^{\dag}_1(\al)
Q(a_1,\cdots a_n,p_1,\cdots ,p_n) \Psi_2(\al)
\label{scaprod}
\fin
where the hermitian conjugate wave function $\Psi^{\dag}_1(\al)$
is defined by:
\debut
\Psi^{\dag}_1(\al)= \bar {\Psi_1({\bar \al})} \label{hermi}
\fin
Notice that although this definition involves two complex
conjugations, it is anti-linear as it should be.
The operator $Q(a_1,\cdots a_n,p_1,\cdots ,p_n)$ inserted
into the integral is not specified here, except for a simple
constraint arising from the condition $\bar {\vev{\Psi_1|\Psi_2}}
= \vev{\Psi_2|\Psi_1}$. The exact form of this
operator has to be fixed from additional requirements.
The role of this operator is similar to that of screening
operators in CFT. This analogy is not just a formal
coincidence as it will be explained later.
Due to commutativity of $A,P$ with $a,p$, the
particular form of the polynomial
$Q$ is irrelevant for formal properties of the
operators $A,P$. The contour $C$ is complicated for
generic $\xi$, but reduces to small circle around the
origin in the reflectionless case.

As usual, given an operator $\CO$, its
hermitic conjugated operator $\CO^{\dag}$ is defined
by $\vev{(\CO^{\dag}\Psi_1)|\Psi_2}=
\vev{\Psi_1|(\CO\Psi_2)}$. For the canonical operators
$A_j$ and $P_j$ this gives:
\debut
A_j^{\dag} &=& A_j \label{APdag}\\
P_j^{\dag} &=& q \prod_{k\not= j}\({\frac{q^2A_j^2-A_k^2}{A_j^2-A_k^2}}\)
\cdot P_j
= q P_j\cdot  \prod_{k\not= j}\({\frac{A_j^2-A_k^2}{ q^{-2}A_j^2-A_k^2}}\)
\fin
It is an interesting check to verify that these relations
are compatible with the Weyl commutation relations (\ref{comrelAP}).
The relation for $A_j^{\dag}$ is obvious from the definition.
The formula for $P_j^{\dag}$ can be deduced as follows: 
\debut
\vev{\Psi_1|(P_j\Psi_2)}&=& \non\\
&& \hskip -2cm
= \int_C d\al_1\cdots\int_C d\al_n~\prod_i A_i
\prod_{i<j}(A^2_i-A^2_j)~ \bar {\Psi_1({\bar \al})}
Q(a_1,\cdots a_n,p_1,\cdots ,p_n)
\Psi_2(\cdots, \al_j +i\xi,\cdots)  \non\\
&& \hskip -2cm
= {\bar q} \int_C d\al_1\cdots\int_C d\al_n~ \prod_i A_i
\prod_{i<j}(A^2_i-A^2_j)~
\prod_{k\not= j}\({\frac{{\bar q}^2A_j^2-A_k^2}{A_j^2-A_k^2}}\)
\bar {\Psi_1(\cdots,{\bar \al_j}+i\xi ,\cdots)}
\non \\
&& \hskip 6cm \times Q(a_1,\cdots a_n,p_1,\cdots ,p_n)\Psi_2(\al)
\hfill \non
\fin
The second equality follows from the first by changing
variables $\al_j\to \al_j+ i\xi$
which is possible if
$\Psi$ is regular as
function of $\alpha _j$ between $C$
and $C+i\xi$. To obtain these  formulae we crucially
use the fact that $q$ is of modulus one, ie. $q^{-1}= \bar q$.
(They will not be correct for real values of $q$.)

The fact that $A_j$ is hermitian simply expresses the fact that the
field $e^{i\varphi}$ is a real field in the RSG theory since it
has to be identified with the field $\Phi_{13}$ of the minimal
conformal model.
In other words, the quantum variables $A_j$ are real
variables leaving on the unit circle!

\subsection{ Hamiltonians and quantum
$\tau$-functions.}

We now introduce the quantum version of the
hamiltonians $H_k$ defined in eq.(\ref{hkclass}). Since the classical
hamiltonians are complicated functions of $A$ and $P$,
we have to specify the order of the operators in the quantum formula.
We choose the following minimal deformation:
\debut
\CT (P'|A)~~  H_k = \CT_k(P'|A) \label{qhamil}
\fin
with $P' = (-1)^n P$ and
\debut
\CT_k(P|A)&=&
\sum_{i_1<i_2\cdots <i_k } \CT (P_1,\cdots ,-P_{i_1}, \cdots,
-P_{i_k}, \cdots ,P_n|A)~ A_{i_1} A_{i_2} \cdots A_{i_k} \label{tauqk}\\
&=& \sum_{i_1<i_2\cdots <i_k } A_{i_1} A_{i_2} \cdots A_{i_k}~
\CT (P_1,\cdots ,-qP_{i_1}, \cdots, -qP_{i_k}, \cdots ,P_n|A) \non
\fin
We shall refer to the operator $\CT (P|A)$ as the `quantum
$\tau$-function'.
Its explicit expression is:
\debut
\CT(P|A) = 1 + \sum_{p=1}^n (-q)^{\frac{p(p-1)}{2}}~
\sum_{{ I \subset \{1,\cdots,n\} \atop |I|=p}}~
(\prod_{{i\in I \atop j\notin I}} \ga_{ij})\cdot \prod_{i\in I} P_i
\label{quantumtau}
\fin
with
\debut
\gamma_{ij} = {q A_i + A_j \over A_i -A_j} \non
\fin
Alternatively, $\CT(P|A)$ can be recursively defined by eq.(\ref{recurq})
in Appendix E.
In the limit $q \to 1$ this formula coincides with
the classical $\tau$-functions $\tau(Z|A)$ as
defined in eq.(\ref{tau1}).  For two particles:
\debut
\CT^{(2)}(P|A) = 1 +{qA_1+A_2 \over A_1-A_2} P_{1}
+ {qA_2+A_1 \over A_2-A_1} P_{2} -q  P_{1}P_{2}
\non
\fin
The operators $\CT(P|A)$, but not the hamiltonians $H_n$,
are closely related to the Mac Donald difference operators \cite{macdo}.
In particular the terms in $\CT(P|A)$ of fixed homogeneity
degree in $P$ form a family of commuting difference operators.
More precisely, the two generating
functions $\CT(uP_1,\cdots,uP_n|A)$ and
$\CT(vP_1,\cdots,vP_n|A)$ commute for any $u$ and $v$.
Note also that $\CT_n(P|A)= \CT(-P|A)\cdot (\prod_iA_i)=
(\prod_iA_i)\CT(-qP|A)$. It is convenient to introduce
the generating function $\hat T(u)$ for the quantum
$\tau$-functions $\CT_k$:
$\hat T(u)= \sum_{k=0}^n~u^k \CT_k$. As shown in Appendix E, we
have the following expression for $\hat T(u)$:
\debut
\hat T(u)= \CT(\hat P(u)|A) \cdot \prod_{j=1}^n(1 + u A_j)
\qquad with \qquad \hat P_j(u) = P_j\cdot \frac{1-uA_j}{1+uA_j}
\label{genequant}
\fin
This can be used to relate the generating functions
of the symmetric functions of the $B$ and $A$ operators as:
\debut
\prod_{j=1}^n(1 + u B_j)\cdot \CT(P|A) =
\CT(\hat P(u)|A) \cdot \prod_{j=1}^n(1 + u A_j) \non
\fin

It is important to realize that all the hamiltonians
$H_k$ commute with the operators $a_j$ and $p_j$, since they
are functions of the $A$ and $P$ only. As we already pointed out,
the operators $a_j$ and $p_j$ are associated to the quantum
affine symmetry $U_{\hat q}(\hat {sl_2})$. Thus, the fact
that the Schr\"odinger equation is a difference equation just
leave enough room for this non-local symmetry.
This symmetry does not have any straightforward classical meaning,
it corresponds to choice of topologically different
components of the classical configuration space
(different cycles). The way of encoding this topological
information into quantum formulae through the algebra which
commute with all local observables is, in our opinion, a very
interesting feature of quantization of solitons.

The conditions for the $H_k$ to be hermitian are bilinear
identities on the quantum
$\tau$-functions :
\debut
\CT(P|A)~ \CT_k(P|A)^{\dag} = \CT_k(P|A)~\CT(P|A)^{\dag}
\label{qbili1}
\fin
Furthermore, the quantum
$\tau$-functions behave nicely under
hermitian conjugation. More precisely,
the definition of $P_j^{\dag}$ implies that
the quantum
$\tau$-functions are not hermitian
for our scalar product but that their hermitian conjugates
are again quantum
$\tau$-functions:
\debut
\CT(\la_1P_1,\cdots,\la_nP_n|A)^{\dag}
= \CT(q\la_1P_1,\cdots,q\la_nP_n|A) \non
\fin
for any real parameters $\la_1,\cdots,\la_n$.
Here, once again we use the fact that $q$ is of modulus one.
In particular, $\CT(P|A)^{\dag}= \CT(qP|A)$.
This allows us to rewrite eq.(\ref{qbili1}) directly in terms
of the quantum
$\tau$-functions.  For example,
for $k=n$ we have $\CT_n(P|A)^{\dag}=
(\prod_iA_i)^{\dag}\CT(-P|A)^{\dag} = (\prod_iA_i)\,\CT(-qP|A)$.
The hermiticity condition of $H_n$ is then equivalent to:
\debut
\CT(P|A)~\CT(-P|A) = \CT(-P|A)~\CT(P|A) \non
\fin
which is a consequence of the identification of the quantum
tau function as the generating function of the Mac Donald
difference operators. Besides the case $k=n$,
we also checked the hermiticity conditions (\ref{qbili1}) for all
two-particle hamiltonians.

The conditions that the hamiltonians $H_k$ commute can also
be rewritten as bilinear identities for the quantum
$\tau$-functions:
\debut
\CT_k(P|A)~\CT_l(P|A)^{\dag} = \CT_l(P|A)~\CT_k(P|A)^{\dag}
\label{hncommute}
\fin
To rewrite these conditions in a simple form,
we used the hermiticity  of the hamiltonians.
The hermiticity condition (\ref{qbili1}) corresponds
to eq.(\ref{hncommute}) with $l=0$.
Eq.(\ref{hncommute}) can be rewritten as a bilinear identity
on the quantum generating function $\hat T(u)$, cf Appendix E,
eq.(\ref{tubili}).
We are missing a complete algebraic proof of these conditions for
arbitrary $n$.  Although we don't have any doubt that it is true
because the hamiltonians admit simultaneous eigenfunctions.

\subsection{Quantum separation of variables and soliton wave functions.}

One of the magic aspect of these quantum hamiltonians is that the
corresponding Schr\"odinger equations admit a separation of variables.
As a consequence, they admit simultaneous eigenfunctions.
Consider the set of $n$ Schr\"odinger equations for a state $\Psi(\al|\b)$:
\debut
\sigma_k(B)\cdot ~\CT (P'|A)\Psi (\al |\b ) = \CT_k (P'|A) \Psi (\al |\b ),
\qquad for\quad k=1,\cdots,n
\label{schrod}
\fin
with eigenvalues $\sigma_k(B)$. (Recall that $P'=(-1)^n P$).
As shown in Appendix E, we can look for  eigenfunctions
$\Psi(\al|\beta)$ in a factorized form:
\debut
\Psi (\al |\b ) = \hat \psi(\al _1|\b ) \hat \psi (\al _2|\b )
\cdots \hat \psi (\al _n|\b )
\label{facto}
\fin
provided the function $\hat \psi(\al |\b )$, which depends only on one of
the $\al$ variables, is solution of the following
separated difference equation:
\debut
P_j\cdot\hat \psi (\al _j|\b ) = \hat \psi(\al _j+i\xi|\b)
= \prod_{k=1}^n \left({B_k-A_j  \over B_k+q A_j } \right)
\hat \psi (\al _j|\b)
\label{quantumPA}
\fin
This is the quantum analogue of the classical separation
of variables. A remarkable fact of eq.(\ref{quantumPA}) is that
its solution can again be factorized into product of functions
depending separately on only one $\b _k$:
$$\hat \psi(\al _j |\b ) =\prod_{k=1}^n \psi(\al _j |\b _k )$$
where $\psi(\al  |\b  )$ satisfies eq.(\ref{eqpsi}).
This probably reflects the
duality symmetry between the $\al$ and $\b$ variables.
It is clear that the function $\psi(\al |\b)$ is defined by
the difference equation up to
multiplication by any $i\xi$-periodic
function. The way to fix this ambiguity is the
following: one requires that the function
$\psi(\al |\b)$ is regular for $0<Im (\al)<2\pi$,
$\psi(\al |\b)=O (\exp ((\pi /\xi -1)\al)$ when $\al\to +\infty$.

\subsection {Exact form factors in the reflectionless case.}

Let us return to the case $\xi =\pi /\nu$ for integer $\nu$.
In this case the wave-functions $\Psi (\al)$ are $2\pi i$-periodic,
that is why we denote them by $\Psi (A)$. Moreover
$\Psi (A)$ are polynomials of $A$, so we take the Hilbert
space as the space of polynomials. The operators $p_j$ are
identically equal to 1 because they correspond to
shift of $\al$'s by  $2\pi i$. Thus the formula for the scalar
product must be of the form
\debut
\vev{\Psi_1|\Psi_2}= \int_C dA _1\cdots\int_C dA _n
\prod_{i<j}(A^2_i-A^2_j)~ \Psi^{\dag}_1(A)
Q(a_1,\cdots a_n) \Psi_2(A)
\non
\fin
In the reflectionless case the contour $C$ are small contour
around the origin. We recall that for $\xi=\frac{\pi}{\nu}$ with
$\nu$ integer the function $\psi(\al|\beta)$ is:
\debut
\psi(\al|\beta)=\prod_{j=1}^{\nu-1}(B- Aq^{-j}) \non
\fin

It is time to discuss the local operators in this $A$-representation.
As we have seen from the exact quantum formulae,
in order to insert the primary field $\Phi _m$
into the matrix element one has to
put under the integral the expression
$$ \prod\limits _{j=1}^n (B_j')^{-{m\over 2}}
\prod\limits _{j=1}^n A_j^m
\prod\limits _{j=1}^n B_j^{-{m\over 2}}$$
On the other hand, in the classical theory
the fields $\Phi_m=e^{im\varphi}$ are represented on the
$n$-soliton solutions by
$\Phi _m = \prod_{j} A_j^m \prod_j B_j ^{-m}$.
After the quantization we have a self-adjoint operator $H_n=H_n(A,P)$
such that $H_n\ \Psi (A,B)=(\prod_j B_j) \Psi (A,B)$.
So, the comparison of the quantum and classical formulae shows that
the classical expression $\prod A_j^m H_n^{-m}$ must be ordered
for quantization as follows
\debut
\Phi _m(A,P)=H_n^{-m/2}(A,P)\left( \prod_{j=1}^n A_j^m \right) H_n^{-m/2}(A,P)
\label{phiquant}
\fin
This ordering prescription ensures that $\Phi _m$ is a real
field, $\Phi _m^{\dag}=\Phi _m$, since $H_n$ and $A_j$ are hermitic.
We hope that the same notation for the operator $\Phi _m$
acting in all the space of states of RSG or restricted to
$n$-soliton subspace is not misleading.

Now we can fix the function $Q(a)$. The contours of integration
are drawn around $A=0$, so $Q(a)$ has to be taken
as $\prod a_k^{-m_k}$: positive powers of $a_k$ would give
zero scalar product. It is also clear that the $m_k$'s must all be different
for the anti-symmetry coming from $\prod _{i<j}(A_i^2-A_j^2)$.
Considering the form of the function $\psi$ one
realizes that if one of the $m_k$ is greater than $n$ then
the matrix element corresponding to the operator $\Phi _1$ vanishes
because the contour of integration with respect to $A_k$ can be
moved to infinity. Thus we are left with the only possible choice
$$m_k=k$$
This is exactly what we have in the formulae known from bootstrap.

\section{Concluding remarks}

Let us describe possible directions of future developments.

We have constructed the local integrals of motion
in terms of the operators $A$ and $P$. One can consider
the descendants of the primary fields with respect to
these operators. However, to consider the
full space of local fields we need to construct
the Virasoro algebras in terms of $A$ and $P$ and
to consider the descendants of the primary fields. We are
sure that it can be done. In this way we must be
able to identify the SG local operators described in
\cite{count} with those coming from the CFT description.
For these computations we do not really need to
study deeper the situation of generic coupling constant
since the formulae from the subsection 5.2 are
absolutely general.

There is another point for which the consideration
of generic coupling constant is important. We have seen
that there is a dual Weyl algebra composed of $a$ and $p$
which commute with the operators $A$ and $P$. It is
easy to argue that the non-local integrals can be
expressed in terms of $a$ and $p$.
The commutativity of local and non-local integrals
follows from the commutativity of $A,P$ with $a,p$.
The non-local charges represent the quantum
loop algebra $U_{\hat{q}}(\hat {sl_2})$.
It would be very interesting to find their expressions
in terms of $a$ and $p$. On the other hand the
non-local charges in the conformal limit correspond to
screening operators.

We want also to remind the reader that we were actually not working
with complete form factors. We omitted certain multipliers
which are the same for all operators, the
normalization of the wave functions in the logic of
this paper. We hope to
explain how this piece appears in our approach in further
publication. Here we just would like to stress
the analogy with the method
of orbits in coadjoint representations. Every orbit is
quantized independently giving an irreducible representation
of the group, but combining them
in the regular representation requires
 to take into account the Plancherel measure: this
is exactly the analog of the omitted normalization factors.

There are amusing coincidences between many tools
used in this paper and those existing in the
works on lattice quantization of SG \cite{fv} :
the Weyl algebras, the functions of the type
of $\psi (\al,\b)$. The latter are now called quantum
dilogarithms because they provide a deformation
of dilogarithm functions.
It might be possible that this coincidence is not
occasional. The classical soliton is similar to
the step-function, so one can imagine that exact
quantum mechanics of $n$-solitons is related to
the theory on the lattice with $n$ sites.

These is a also a no-empty intersection with recent works on
the Calogero models, Mac Donald polynomials and affine
Hecke algebras. We realize that techniques very similar to
those we used to deal with the quantum
$\tau$-function, i.e.
the generating function of Mac Donald difference operators,
may be used to separate the variables in these operators.
One potential application of this remark would be a more
detailed description of the algebra of the Mac Donald polynomials.

\vfill \eject

\section{Appendix A: The analytical variables and finite zone solutions.}

Let us explain the origin of the $A$ variables as a remnant of
the parametrization of the finite zone solutions, when they degenerate
to the soliton case. We shall do it in the simplest case of the KdV
equation, cf eg. \cite{novbook}.  We start with an hyperelliptic curve
 $\Gamma$ of genus $n$, and a divisor $D$ of degree $n$ on it.
\debut
\Gamma : s^2 &=& R(\lambda),~~ R(\lambda) = \prod_{j=0}^{2n}(\lambda
-\lambda_j) \non \\
D : &=& (\nu_1, \nu_2, \cdots, \nu_n ) \non
\fin
We describe $\Gamma$ as a two sheeted cover of the
$\lambda$ plane. We put cuts on the real axis on the intervals
 $(-\infty, \lambda_0)$ and $(\lambda_{2i-1}, \lambda_{2i})$,
 $i=1,\cdots, n$. The quantities $\nu_i$ in the divisor $D$ denote the $\lambda$
coordinates of the points of the divisor. One should keep in mind that to specify the
points themselves, one has to choose the sheet above $\lambda =\nu_i$.
With these data we construct the Baker-Akhiezer function which is the unique
function with the following analytical properties:
\begin{itemize}
\item It has an essential singularity at the point $P_+$ above infinity:
$\psi(x,\lambda) = e^{k x}(1 + O(1/k))$ with $k = \sqrt{ \lambda }$.
\item It has $n$ simple poles outside $P_+$. The divisor of these
poles is $D$.
\end{itemize}
Considering the quantity $-\partial_x^2 \psi + \lambda \psi$, we see that
it has the same analytical properties as $\psi$ itself, apart for the first 
normalization condition. Hence, because $\psi$ is unique,
 there exists a function $u(x)$ such that
\debut
-\partial_x^2\psi + u(x) \psi + \lambda \psi =0
\label{A0}
\fin
We recognize the usual linear system associated to the KdV equation.
One can give various explicit constructions of the Baker-Akhiezer
 function. The most
popular one is in terms of theta functions. However, for our purpose,
another representation is more suitable.
Let us introduce the divisor  $Z(x)$  of the zeroes of the Baker-Akhiezer function. It
is of degree $n$:
\debut
Z(x) : = (\mu_1(x), \mu_2(x), \cdots, \mu_n(x) ) \non
\fin
One can find the
equations of motion for the divisor $Z(x)$. Consider the function
$\partial_x \psi / \psi $. It is a meromorphic function on $\Gamma$, it
has poles at the points $\mu_i(x)$ and behaves like $ k + O(1/k)$ in the
vicinity of the point $P_+$. Hence we can write
\debut
{\partial_x \psi \over \psi} =  {\sqrt{ R(\lambda)} + Q(x,\lambda)
\over \prod_{i=1}^g (\lambda - \mu_i(x) ) }
\label{A1}
\fin
where $Q$ is a polynomial of degree $n-1$ in $\lambda$. We determine $Q$ by requiring
that $\partial_x \psi \over \psi$ has a pole above $\lambda = \mu_i(x)$
only on one of the two sheets (say $ \sqrt{ R(\lambda)}$). Then
\debut
Q(x,\mu_i(x)) = \sqrt{ R(\mu_i(x))}
\label{A2}
\fin
Hence
\debut
Q(x,\lambda) =\sum_i \sqrt{ R(\mu_i(x))} {\prod_{j \neq i}(\lambda -
\mu_j(x)) \over \prod_{j \neq i} (\mu_i(x) - \mu_j(x) ) }
\label{A3}
\fin
On the other side, in the vicinity of $\mu_i(x)$, we have:
\debut
{\partial_x \psi \over \psi} = - {\partial_x \mu_i(x) \over \lambda -
\mu_i(x) } + O(1)
\label{A4}
\fin
Comparing eq.(\ref{A1}) and eq.(\ref{A4}), we get the equation of motions:
\debut
\partial_x \mu_i(x) = -2 {  \sqrt{ R(\mu_i(x))} \over \prod_{j \neq i}
(\mu_i(x) - \mu_j(x) ) }
\label{A5}
\fin
One can now reconstruct the Baker-Akhiezer function itself. Indeed,
inserting eq.(\ref{A5}) into (\ref{A1}) we get:
\debut
{\partial_x \psi \over \psi} =  {\sqrt{ R(\lambda) } \over P(\lambda,x)}
-{1\over 2} \sum_i {1\over \lambda -\mu_i(x) } \partial_x \mu_i(x)
\non
\fin
where the polynomial $P(\lambda,x)$ is defined as $P(\lambda,x) = \prod_i (\lambda
-\mu_i(x))$.  Therefore \cite{novbook}:
\debut
{\psi(\lambda,x)\over \psi(\lambda,x_0)} = 
\sqrt{ P(\lambda,x) \over P(\lambda, x_0) }
\exp \left( { \int_{x_0}^x { \sqrt{R(\lambda)} \over P(\lambda, x)} dx }\right)
\label{A6}
\fin

One can also reconstruct the potential $u(x)$ directly in terms of the data
$\mu_i(x)$ and $\lambda_j$. Inserting back eq(\ref{A6}) into eq.(\ref{A0}),
we get the polynomial identity
\debut
R = - {1\over 2} P P'' + {1\over 4} P'^2 + (u+\lambda)P^2
\non
\fin
Comparing  the terms $\lambda^{2n}$ we obtain:
\debut
u = 2 \sum_{i=1}^n \mu_i(x) - \sum_{i=0}^{2n} \lambda_j
\label{uzone}
\fin
We are now ready to analyze the soliton limit.
It corresponds to the following limiting configuration:
\debut
\lambda_0 =0, ~~~\lambda_{2j-1}= \lambda_{2j} = B_{j}^{2};~~~
 \lambda = A^{2};~~~ \mu_j = A_{j}^{2}
\label{degeneration}
\fin
In this limiting configuration, eq.(\ref{uzone}) reduces to eq.(\ref{ua}).
Under the condition (\ref{degeneration}), the full curve $\Gamma$
becomes the $A$ plane but with the points $B_j$ and $-B_j$, $j =
1,\cdots,n$ identified. In this limit we have:
\debut
\sqrt{R(\lambda)}\Big\vert_{\la=A^{2}} =
- A \prod_{j=1}^n \left(  A^2 - B_j^2 \right),~~~
P(\lambda,x)\Big\vert_{\la=A^{2}} = \prod_{j=1}^n
\left({A^2 -A_j^2  }\right)
\non
\fin
The equations of motion for the $A_i$ become (compare with
eq.(\ref{eqa1}))
\debut
\partial_{x} A_i =  \prod_j { (A_i^2 - B_j^2 )}
 \prod_{j \neq i}{ 1 \over A_i^2 -A_j^2} \label{motion3}
\fin
We can also obtain the degeneration of the Baker-Akhiezer function. Noticing
that
\debut
{ \sqrt{R(\lambda)} \over P(\lambda,x)}\Big\vert_{\la=A^{2}} =
- A -{1\over 2}\sum_i \left( {1\over A-A_i} + {1\over
A+A_i} \right) \partial_{x} A_i
\non
\fin
we can integrate eq.(\ref{A6}) to get
\debut
{\psi(A,x)\over \psi(A,x_0)} = \prod_{j=1}^n \(
{A - A_j(x) \over A - A_j(x_0)  }\) ~\exp
 \left(- A(x -x_0) \right)
\label{bakersol}
\fin
As we see, in this limit the Baker-Akhiezer function contains a single
 exponential factor instead
of two as one would expect from the second order linear equation
eq.(\ref{A0}). This correspond exactly to the fact that for soliton
solutions the potential $u$ in eq.(\ref{A0}) is reflectionless.
It is manifest in eq.(\ref{bakersol}) that the variables $A_j$ are
the zeroes of the Baker-Akhiezer function.

As  we said, the points $B_j$ and $-B_j$ are identified. This means
that the Baker-Akhiezer function should satisfy the conditions $\psi (B_i,x) =
\psi(-B_i,x)$, $i= 1,\cdots n$. Writing theses conditions, we get
\debut
\prod_{j=1}^n {B_i -A_j(x) \over B_i + A_j(x) } = Y_i e^{ 2 B_i x}
\label{solumotion}
\fin
where $Y_i$ depends only  on the initial conditions at $x=x_0$.
 Comparing with eqs.(\ref{evolution},\ref{XtoA}),
we recognize the solution of the equations of motion.

In the sine-Gordon soliton case, we identify $x \equiv x_-$. 
The $x_+$ dependence is simply reintroduced 
by replacing in eq.(\ref{solumotion})
$$ e^{ 2 B_i x} \to e^{ 2 B_i x_- + 2 B_i^{-1} x_+} $$

Let us discuss now the motion of the divisor $Z(x_-,x_+)$. In the finite
zone case, the point $\mu_i(x_-)$ has a quasi-periodic motion
 on the real axis in the interval
$\lambda_{2i-2} \leq \mu_i(x_-) \leq \lambda_{2i-1}$.

Consider the $x_+$ motion in the soliton case when we have only
two points. When $x_+=-\infty$, the points $A_1$ and $A_2$ start from
$B_1$ and $B_2$ respectively. When $x_+$ increases, the points $A_1$ and $A_2$
start to move to the right. Notice that the point $A =\infty$ is
regular for eq.(\ref{motion3}), so the point $A_2$ passes smoothly
from $+\infty$ to $-\infty$, and then continues to move towards
$-B_2$ which is reached at some time $x_*$. At the {\it same} time $x_*$,
$A_1$ reaches $B_2$ so that in the right hand side of
eq.(\ref{solumotion}) the pole at $A_2=-B_2$ is cancelled by the zero at
$A_1 = B_2$. Hence everything remains finite for a finite time $x_*$.
 At this time, the point $A_1$ jumps to $-B_2$ and
ends  its motion at the point $-B_1$. Similarly, $A_2$ jumps to $B_2$
and continues its motion up to $-B_2$ again. The case of generic $n$
is similar. Altogether, $A_i$ starts at $B_i$ and ends at $-B_i$.

Eq.(\ref{bakersol}) is easily related to the Jost solution of eq.(\ref{A0}).
In our context, the Jost solution is defined by the normalization condition
$\lim_{x_- \to - \infty} \psi_{Jost}(x_-) = \exp (-A x_-)$. Hence, we find
\debut
\psi_{Jost}(x_-) = \lim_{x_0 \to - \infty} e^{-A x_0} {\psi_{Baker}(x_-)\over
\psi_{Baker}(x_0)}
 = \prod_{j=1}^n \left( {A- A_j(x_-)  \over A- B_j } \right)
 \exp \left(-Ax_-\right)
\non
\fin
where we used the fact that $\lim_{x_0 \to - \infty} A_j(x_0) = B_j$.
Therefore $B_j$ are the poles of the Jost solution and $A_j(x)$ are its
zeroes.

Finally, we would like to discuss the important question of
the Poisson structures of the KdV equation.
As we know, we have a whole hierarchy of these structures.
One can describe the restrictions of the symplectic forms to
the manifold of finite zone solutions \cite{novbook}. Let $\Omega^{(k)}$ be the restricted
$k$-th symplectic form. Then we have in terms of analytical variables
\cite{fm,nv}:
\debut
\Omega^{(k)} = \sum_{i=1}^n d {\cal P}(\mu_i)\wedge {d \mu_i \over
\mu_i^{k-1}}
\label{novom}
\fin
where ${\cal P}(\lambda)$ is the pseudo momentum  defined by:
\debut
{\cal P}(\lambda) = \log {\psi (\lambda, x=L) \over \psi(\lambda,x=-L)}
\non
\fin

To compute the quasi-momentum in the soliton limit, we choose the normalization point to be
$x_0= -L$, and we send $L \to \infty$. According to the previous discussion,
 we have $A_i(L) \to -B_i$, $A_i(-L) \to B_i$. Using eq.(\ref{bakersol}), we get
\debut
{\cal P}(A) =- { 2 L  A} - \log \prod_j\({B_j-A \over B_j+A}\)~~~ mod~ ~i\pi
\non
\fin
Hence,
\debut
 \Omega^{(k)} =2\sum_{i=1}^n  d \log \prod_j \({B_j-A_i \over B_j+A_i}\)
\wedge {dA_i \over A_i^{2k-3}}
\non
\fin
The form used in eq.(\ref{1fm}) corresponds to $k=2$ i.e., to the second
Hamiltonian structure of KdV. We recall once more that the second
Hamiltonian structure of the KdV equation is precisely the Virasoro algebra.

\section{Appendix B: From the $\{X,B\}$ to the $\{A,B\}$ variables.}

Before explaining the proof of the formula for the $\tau$-functions
in the $\{A,B\}$ variables we need to gather a few facts
concerning the $\tau$-functions.
The $\tau$-functions are defined by the determinant (\ref{deftau}):
$\tau_\pm= \det(1\pm V)$. Most of the proof will be recursive
using the recursion relation (\ref{recur1}) satisfied by them,
cf eg. ref.\cite{bb2}:
\debut
\tau ^{(n)}(X|B)= \tau ^{(n-1)}(X|B) +
\tau^{(n-1)}(\beta_{kn}^2(B)X_k|B)~X_n.
\label{recur2}
\fin
For comparison with the quantum formula, it is also useful to know
the explicit expression of the $\tau$-function not in the $X_j$
variables but in the variables
$Y_j=X_j \prod_{k\not= j}\(\frac{B_j-B_k}{B_j+B_k}\)$:
\debut
\tau^{(n)}(Y|B)  = 1 + \sum_{p=1}^n
\sum_{{ I \subset \{1,\cdots,n\} \atop |I|=p}}
\prod_{{i \in I \atop j\notin I}} \beta^{-1}_{ij}(B)\cdot \prod_{i\in I} Y_i
\label{tauY}
\fin
with $\beta_{ij}(B)=\frac{B_i-B_j}{B_i+B_j}$.

Let us now prove the formula (\ref{tauAB1}) for the $\tau$-functions.
We recall them to ease the reading:
\debut
\tau_+^{(n)} &=& 2^n \left( \prod_{j=1}^n B_j\right)
 { \prod_{i < j} (A_i +A_j) \prod_{i <j} (B_i +B_j)
\over \prod_{i,j} (B_i + A_j) }
\label{tauAB} \\
\tau_-^{(n)} &=& 2^n \left( \prod_{j=1}^n A_j\right)
 { \prod_{i < j} (A_i +A_j) \prod_{i <j} (B_i +B_j)
\over \prod_{i,j} (B_i + A_j) }
\non
\fin
The upper index $n$ refers to the $n$-soliton solutions.
Eqs.(\ref{tauAB}) are
two identities between rational functions of $A$ and $B$
once the expression of $X$ as a function of $A$ and $B$,
\debut
Y_j^{(n)}= X_j^{(n)} \cdot \prod_{k \neq j}\left( {B_j-B_k \over B_j+B_k}\right)
= \prod_{k=1}^n \left({ B_j -  A_k \over B_j + A_k } \right)
\label{XtoAbis}
\fin
has been inserted into the $\tau$-functions.

Let us denote by $\hat \tau_\pm^{(n)}(A|B)$ the $\tau$-functions
with $X_j$ expressed in terms of $A$ and $B$.
Since $\tau_\pm^{(n)}(X|B)$ are symmetric in $X_j$, so is
$\hat \tau_\pm^{(n)}(A|B)$ as a function of $A_j$.
Since permuting the $B$ permutes the $X$, the
functions $\hat \tau_\pm^{(n)}(A|B)$ are also symmetric
in the $B$. Thus, the identities (\ref{tauAB}) are equalities
between rational functions symmetric in $A$ and $B$.

Let us first show that $\hat \tau_\pm^{(n)}(A|B)$ has poles
only at $A_j+B_k=0$. In view of the explicit expressions
 of the $\tau$-functions and of the $X_j^{(n)}$,
$\hat \tau_\pm^{(n)}(A|B)$ has potentially
simple poles at $A_j+B_k=0$ and $B_j\pm B_k=0$.
The expression (\ref{tauY}) of the $\tau$-function in
terms of $Y$ shows that there are no poles at $B_j+B_k=0$.
Similarly, the expression (\ref{tau1}) of the $\tau$-functions
in terms of $X$ shows that the potential poles at $B_j-B_k=0$ are
associated to $X_j$ and $X_k$. Using twice the recursion
relation (\ref{recur2}) shows that these poles cancel against
$\beta^2_{ij}(B)$.  Thus, $\hat \tau_\pm^{(n)}(A|B)$ can be written as
\debut
\hat \tau_\pm^{(n)}(A|B) =
\frac{ Q_\pm^{(n)}(A|B) }{\prod_{j,k}(A_j+B_k)}
\non
\fin
where $Q_\pm^{(n)}(A|B)$ are polynomials, symmetric in $A$ and $B$,
and of degree at most $n$ in  each variables.

To prove the identities (\ref{tauAB}) it is then  enough
by symmetry to show that the functions have identical residues and
that they coincides at  particular points.
Again by symmetry, it is enough to check it for
the pole at $A_n+B_n=0$ and at the point $A_n=B_n$.
To keep the size of this Appendix reasonable we shall
describe the proof for $\tau_+$ only. The proof
for $\tau_-$ is similar.
We shall prove it by induction assuming that the
identities are true up to $n-1$ solitons (they are
obviously true in the one soliton case).
The recursive proof is based on the three following relations:

\noindent i) The $\tau$-function satisfies the recursion
relation (\ref{recur2}).

\noindent ii) The r.h.s of eq.(\ref{tauAB})
satisfies the following recursion relation:
\debut
rhs^{(n)} = \({\frac{2B_n}{A_n+B_n}}\) \prod_{j\not= n}
\frac{(A_n+A_j)(B_n+B_j)}{(A_n+B_j)(B_n+A_j)}~\cdot~ rhs^{(n-1)}
\label{rhs1}
\fin

\noindent iii) As functions of $A$ and $B$ the variables $X_j$
satisfy the following recursion relations:
\debut
X_j^{(n)}&=& \({\frac{B_n+B_j}{B_j-B_n}}\)
\({\frac{B_j-A_n}{B_j+A_n}}\)~ X_j^{(n-1)}
\qquad for \quad j=1,\cdots, n-1
\label{recurX1} \\
X_n^{(n)}&=& \({\frac{B_n-A_n}{B_n+A_n}}\)
\prod_{j\not= n}\({\frac{B_n+B_j}{B_n-B_j}}\)
\({\frac{B_n-A_j}{B_n+A_j}}\)
\label{recurX2}
\fin
where $X_j^{(n-1)}$ is independent of $A_n,B_n$.

Let us first compare the values of both side of
 eq.(\ref{tauAB}) at the point $A_n=B_n$.
Using eq.(\ref{rhs1}), we have for the r.h.s.:
\debut
rhs^{(n)}\Big\vert_{A_n=B_n} = rhs^{(n-1)}
\non
\fin
For the other side we remark that $X_n^{(n)}$ vanishes
for $A_n=B_n$. Therefore the recursion relation
(\ref{recur2}) for the $\tau$-functions implies that:
\debut
lhs^{(n)}\Big\vert_{A_n=B_n} = lhs^{(n-1)} \non
\fin
Thus, both sides of eq.(\ref{tauAB}) coincide at $A_n=B_n$.

Let us now compute the residue at the simple pole $A_n=-B_n$.
The pole in $\hat \tau_+^{(n)}(A|B)$ comes from the pole
of $X_n^{(n)}$ at $A_n=-B_n$. Its residue is:
\debut
Res(X_n^{(n)})\Big\vert_{A_n=-B_n}= 2B_n
\prod_{j\not= n}\({\frac{B_n+B_j}{B_n-B_j}}\)
\({\frac{B_n-A_j}{B_n+A_j} }\)
\non
\fin
Furthermore, eq.(\ref{recurX1}) gives for $1\leq j\leq n-1$:
\debut
X_j^{(n)}\Big\vert_{A_n=-B_n} = \beta_{jn}^{-2}(B)\, X_j^{(n-1)}
\non
\fin
Therefore the factor $\beta^2_{jn}(B)$ cancels in
the recursion relation (\ref{recur2}) and we get:
\debut
Res(\hat \tau_+^{(n)}(A|B))\Big\vert_{A_n=-B_n}
= Res(X_n^{(n)})\Big\vert_{A_n=-B_n}~\cdot~
\hat \tau_+^{(n-1)}(A|B)) \non
\fin
On the other hand, eq.(\ref{rhs1}) implies:
\debut
Res(rhs^{(n)})\Big\vert_{A_n=-B_n}&=& 2B_n \prod_{j\not= n}
\frac{(A_j-B_n)(B_n+B_j)}{(B_j-B_n)(B_n+A_j)}~\cdot~ rhs^{(n-1)}\non\\
&=& Res(X_n^{(n)})\Big\vert_{A_n=-B_n}~\cdot~ rhs^{(n-1)} \non
\fin
Thus, the residues of both sides of eq.(\ref{tauAB}) at $A_n=-B_n$
coincide by the induction hypothesis.
This concludes the proof of the identities (\ref{tauAB}).

Let us now consider the formula expressing the symmetric
functions $\sig_k(A)$ as functions of the $\{X,B\}$ variables.
As pointed out in the main text, the defining relations
(\ref{XtoA}) or (\ref{XtoAbis}) can be considered
as a system of equations for the symmetric functions of $A$.
We claim that the solution to this system can be written as:
\debut
\sigma _k(A)={\tau _k(X'|B)\over \tau (X'|B)}
\label{solA}
\fin
where $X'= (-1)^n X$ and
\debut
 \tau _k(X_1,\cdots ,X_n|B_1,\cdots ,B_n)&=&  \non \\
&&\hskip -3cm
\sum\limits _{i_1<i_2<\cdots <i_k} B_{i_1}B_{i_2}\cdots B_{i_k}
\tau(X_1,\cdots ,-X_{i_1},\cdots,-X_{i_k} ,\cdots,X_n|B_1,\cdots ,B_n)
\non
\fin
Let us compare eqs.(\ref{XtoA}) and
(\ref{defP}). A quick look reveals that they are identical provided
we exchange $A$ with $B$ and $Y$ with $P$. Hence, the proof of
eq.(\ref{solA}) is identical to the proof of eq.(\ref{hkclass}),
which is a limiting case of the quantum formula whose proof is
given in Appendix E.

This somewhat mysterious formula can be related to well known results
connecting the Baker function and the tau function.
One first reintroduce all the times by the substitution
$ X_i \to X_i(t)= X_i(0)\exp( 2 \xi(B_i,t)) $,
where $ \xi(A,t) = \sum_i A^{2i-1}t_{2i-1}$.
Next, using the generating function eq.(\ref{tuclass}) we have
\begin{eqnarray}
\prod_i(1+u A_i)&=& \sum_k u^k {\tau_k \over \tau}=
\prod_j (1+u B_j){\tau( {1-uB_i\over 1+u B_i}X_i|B) \over 
\tau(X|B) } \nonumber
\end{eqnarray}
On the other hand, the multi time Baker function reads
\begin{eqnarray}
{\psi(A,t) \over \psi(A,t^{(0)}) } &=& 
\prod_i \left( {1-A_i(t)/A \over 1- A_i(t^{(0)})/A} \right) 
e^{-\xi(A,t)+ \xi(A,t^{(0)})}
\nonumber
\end{eqnarray}
Combining these two 
formulae, we can identify the Baker function as
\begin{eqnarray}
\psi(A,t) &=& {\tau( {1+B_i/A\over 1- B_i/A}X_i|B) \over 
\tau(X|B) } e^{-\xi(A,t) }
\nonumber
\end{eqnarray}
Since 
${1+B_i/A\over 1- B_i/A} = \exp\left( 2 \sum_{i} {1\over 2n-1} \left(
{B_i\over A}\right)^{2n-1} \right)$
we find
\begin{eqnarray}
\psi(A,t) &=& {\tau(t_{2n-1} +{1\over 2n-1} 
{1\over A^{2n-1}}) \over 
\tau(t) } e^{-\xi(A,t) }
\nonumber
\end{eqnarray}
and we recognize the well known Sato formula.

\section{Appendix C: Information about the integral formulae.}

In this appendix we explain how the formulae for the form  factors
given in the Section 3 agrees with the conventional ones \cite{book}.
We have to show how to reduce the number of integrals
and the degree of polynomials.

In order to reduce the number of integrals by one let us
consider the integral with respect to $A_n$ in which we would
like to move the integration contour to infinity.
When $A_n\to\infty$ one has
\debut
\prod\limits _{j=1} ^{2n} \psi (A_n,B_j)
\prod\limits _{i<n} (A_i^2-A_n^2) a_n^{-n}=
A_n^{-2}\left( 1-{q+1\over q-1}A_n^{-1}\sum_j B_j +O(A_n^{-2})\right)
\label{inf}
\fin
So, the integral over $A_n$ is
\debut
{1\over 2\pi i}\int\limits _C
A_n^{-2}\left(1-{q+1\over q-1}A_n^{-1}\sum_j B_j +O(A_n^{-2})\right)
L_\CO ( A_1,\cdots ,A_{n-1},A_n|B_1,\cdots ,B_{2n})\ dA_n \label{An}
\fin
It is important that there are no contributions with $a_n=A_n^{2\nu}$
which means that the integral is essentially independent of the
coupling constant (it depends on $\nu$ only through the constants
like ${q+1\over q-1} $ in (\ref{An})).
The first two terms written in (\ref{An}) are sufficient to
calculate this integral for
$$ L_{\Phi _1}(A_1,\cdots ,A_n|B_1,\cdots ,B_{2n})
=\prod_jA_j \prod_j B_j^{-{1\over 2}} $$
and
$$ L_{\Phi _2}(A_1,\cdots ,A_n|B_1,\cdots ,B_{2n})
=\prod_j A_j^2 \prod_j B_j^{-1} $$
the results being respectively
$$ \prod\limits_{i=1}^{n-1} A_i \prod_j B_j^{-{1\over 2}}
\qquad and \qquad
{q+1\over q-1}(\sum_j B_j)\prod\limits_{i=1}^{n-1} A_i^2 \prod_j
B_j^{-1}  $$
This calculation gives agreement with the formulae from \cite{book}.
To calculate the integral over $A_n$ for higher operator $\Phi _m$
one has to take into account higher contributions to (\ref{inf}),
it would be nice to find these primary operators among those
described in \cite{count}.

Let us now explain why the reduction of the degree of polynomials
is possible.
The formula for $\widehat{f}_\CO $ is composed of one-fold
integrals of the type
$$
\int\limits _C
\prod\limits _{j=1} ^{2n} \psi (A,B_j)
a^{-k} L(A) dA $$
One can reduce the degree of the polynomial $L(A)$ using the following
circumstance.
Due to the fact that the function $\psi (A,B)$
satisfies the equation (\ref{eqpsi})
$$ \psi(Aq,B)={B-A\over B+qA} \psi(A,B)$$
the polynomials of the form
\debut
L(A)=M(A)\prod_j (B_j+A)-qM(Aq)\prod_j (B_j-A) \simeq 0
\label{zero2}
\fin
for any polynomial $M(A)$.  Here $L(A)\simeq 0$ means that $L(A)$
produces zero when substituted into the integral. This allows
for any given polynomial $L(A)$ to find such a polynomial $L'(A)$
such that $L(A)\simeq L'(A)$ and $deg(L')\le 2n-1$. To do that
one has to find a polynomial $M(A)$ such that
$$L(A)=M(A)\prod_j (B_j+A)-qM(Aq)\prod_j (B_j-A) +L'(A)$$
with $deg(L')\le 2n-1$. This is always possible by induction.

\section{Appendix D: A semi-classical limit.}

We start with
\debut
\psi (A,B) = \prod_{j=1}^{\nu -1} (B- Aq^{-j}); ~~~~ q=e^{i{\pi \over \nu}}
\non
\fin
We recall the obvious formula
\debut
\sum_{n=1}^{N-1} f(n \Delta )&=& {1\over \Delta}\sum_{n=0}^{N-1}
{f(n\Delta) + f((n+1)\Delta) \over 2}\Delta -{1\over 2}(f(0)+ f(N))
\non \\
&=& {1\over \Delta} \int_0^{N\Delta} f(x) dx - {1\over 2}(f(0)+ f(N))
+O(\Delta)
\non
\fin
Using this formula we find
\debut
\log \psi &=& {\nu \over \pi } \int_{0}^\pi \log (B-A e^{-ix} ) dx
- {1\over 2}( \log(B-A ) + \log( B+A)) +O(\Delta)
\non
\fin
The integral defines an analytical function of $A$ in the plane with a cut
which is a semi circle from $B$ to $-B$ in the upper half plane. Remark that
for $A=0$, this integral is real and its value is $\pi \log B$. To proceed, we perform
some formal manipulations on the integral. We have
\debut
\int_{0}^\pi \log (B-A e^{-ix} ) dx
&=& i \int_{-A}^A \log( B+A){dA\over A} =
i\int_{0}^A \log( B+A){dA\over A}
  -i \int_{0}^{-A} \log( B+A){dA\over A} \non \\
&=&i \int_{0}^A \log\({B+A \over B-A}\) {dA\over A} + \pi \log B
\non
\fin
where the last term has been added to  normalize the function by its value at
$A=0$, thereby fixing the ambiguities of the formal manipulations.
 Putting everything together, we get
\debut
\psi(A,B) = { B^\nu \over  \sqrt{B^2 -A^2}} ~
\exp \left[{-i{\nu \over \pi }\int_{0}^A \log \left({ B-A \over
B+A}\right) {dA\over A} } \right]
\non
\fin
which is the result quoted in the text.

\section{Appendix E: A proof of the quantum separation of variables.}

Before proving the separation of variables in the quantum theory,
we present a proof for the formula of the generating function
$\hat T^{(n)}(u)= \sum_k u^k \CT_k^{(n)}$.
As it is defined in (\ref{quantumtau}), the quantum
$\tau$-function
satisfies the following recursion relation:
\debut
\CT^{(n)}(P|A)= : \CT^{(n-1)} ( \gamma_{kn}P_k|A_k ) +
( \prod_{k \neq n} \gamma_{nk} )\, \CT^{(n-1)}
\left(\overline{\gamma}_{kn}^{-1} P_k|A_k\right)\, P_n :
\label{recurq}
\fin
where
\debut
\gamma_{ij}={q A_i + A_j \over A_i -A_j},\qquad and\qquad
\overline{\gamma}_{ij}= {q^{-1}A_i + A_j \over A_i - A_j }= -{\bar q}\ga_{ji}
\non
\fin
The double dots $:~:$ mean writing the $P$'s on the right.
In the classical limit $q\to 1$ this is equivalent to the
relation (\ref{recur1}).

Let $\CT_k^{(n)}$ be the operators defined in eq.(\ref{tauqk})
in the $n$ solitons case.  By convention we set:
$\CT_0^{(n)}= \CT^{(n)}(P|A)$ and $\CT_k^{(n)}=0$ for $k<0$ or $k>n$.
From eq.(\ref{recurq}), we deduce a recursion relation for the $\CT_k$:
\debut
\CT_k^{(n)} = \tilde \CT_k^{(n-1)} + \tilde {\tilde \CT}_k^{(n-1)}\ P_n
+A_n\({ \tilde \CT_{k-1}^{(n-1)} - q \tilde {\tilde \CT}_{k-1}^{(n-1)}\ P_n}\)
\label{recurtk}
\fin
where
\debut
\tilde \CT_k^{(n-1)} &=& :\CT_k^{(n-1)}(\gamma_{jn}P_j|A_j ): \non\\
\tilde {\tilde \CT}_k^{(n-1)}&=&
:( \prod_{j \neq n} \gamma_{nj} )\, \CT_k^{(n-1)}
\left(\overline{\gamma}_{jn}^{-1} P_j|A_j\right) :\non
\fin
Summing up eqs.(\ref{recurtk})
and defining $\tilde T^{(n-1)}(u)=\sum_k u^k \tilde \CT_k^{(n-1)}$
and $\tilde {\tilde T}^{(n-1)}(u) = \sum_k u^k \tilde {\tilde \CT}_k^{(n-1)}$,
we get:
\debut
\hat T^{(n)}(u) &=& (1+uA_n) \tilde T^{(n-1)}(u) +
(1-quA_n)\tilde {\tilde T}^{(n-1)}(u) P_n \non\\
&=& \({ \tilde T^{(n-1)}(u) + \tilde {\tilde T}^{(n-1)}(u)
 \ P_n\({\frac{1-uA_n}{1+uA_n} }\)}\) (1+uA_n) \non
\fin
Comparing with the recursion relation (\ref{recurq})
satisfied by the quantum
$\tau$-function $\CT(P|A)$ proves the result
quoted in eq.(\ref{genequant}).

Furthermore, the hermiticity properties of the
quantum
$\tau$-function $\tau(P|A)$ implies:
\debut
\hat T(u)^{\dag} =  \prod_{j=1}^n(1+uA_j)
\CT(q\hat {\hat P}(u)|A), \qquad with \qquad
\hat {\hat P}_j(u)= \frac{1-uA_j}{1+uA_j} P_j \non
\fin

We can use the generating function $\hat T(u)$ do write
the commutativity relation (\ref{hncommute}) in an
alternative form:
\debut
\CT(\hat P(u)|A)\cdot \prod_{j=1}^n(1+uA_j)
(1+vA_j)\cdot \CT(q\hat {\hat P}(v)|A)
= \CT(\hat P(v)|A)\cdot \prod_{j=1}^n(1+vA_j)
(1+uA_j) \cdot\CT(q\hat {\hat P}(u)|A) \label{tubili}
\fin

Let us now give a proof of the quantum separation of variables.
Assume that we are acting on the quantum hamiltonian $H_k$
with a wave function $\Psi(\al|\beta)$, eq.(\ref{facto}),
 satisfying the difference equation (\ref{quantumPA}).
Since the momenta operators $P_j$
have been ordered to the right in the definition
of the quantum
$\tau$-functions, this wave function will be an eigenfunction with
eigenvalue $\sig_k(B)$ if the following relation is true:
\debut
\sig_k(B)~~\CT^{(n)} (P^{(n)}|A)  = \CT^{(n)}_k(P^{(n)}|A)
\label{tobe}
\fin
with
\debut
P^{(n)}_j= \prod_{k=1}^n\({\frac{A_j-B_k}{qA_j+B_k} }\)
\label{pn}
\fin
Recall that we define the Hamiltonians using $P'= (-1)^nP$.
In eq.(\ref{tobe}) we used the same notation for the
quantum operator $\CT^{(n)}(P'|A)$ and the c-number function
obtained by inserting the values (\ref{pn}) of the
momenta operators $P'_j$.
We recall that the functions $\CT^{(n)}_k(P|A)$ are defined
by:
\debut
\CT^{(n)}_k(P|A)=
\sum_{i_1<i_2\cdots <i_k }\  A_{i_1} A_{i_2} \cdots A_{i_k}\
\tau (P_{1},\cdots ,-qP_{i_1}, \cdots,
-qP_{i_k}, \cdots ,P_n|A)
\label{tqkbis}
\fin

Once the specific values (\ref{pn}) of the momenta
have been plugged into eq.(\ref{tobe}), this is an identity between two
rational functions in the variables $A$ and $B$.
Both the l.h.s. and the r.h.s. of (\ref{tobe})
 are symmetric functions in $A$ and $B$.
They only have simple poles at $qA_j+B_k=0$ and $A_j=A_k$,
the former comes from $P^{(n)}_j$ and the latter from $\ga_{jk}$.
Due to their behavior at infinity,  to prove eq.(\ref{tobe})
it is enough to check that these rational functions
 have the same residues at
their poles and that they are equal at specific points.
By symmetry it is enough to check the residues at
$qA_n+B_n=0$ and $A_n=A_{n-1}$, and the values at
the point $A_n=B_n$. We shall do it by induction
assuming that the identity (\ref{tobe}) is true up
to $n-1$ (it is of course true for $n=1$).

i) Consider first the residue at $qA_n+B_n=0$.
This pole is associated to $P^{(n)}_n$. To compute the
residues of both side of eq.(\ref{tobe}) we may use
the recursion relation (\ref{recurq}).
Using the fact that:
\debut
\inv{\bar \ga_{jn}} P^{(n)}_j\Big \vert_{qA_n+B_n=0} = P^{(n-1)}_j,
\qquad for\qquad j=1,\cdots,n-1
\non
\fin
the recursion relation (\ref{tobe}) implies:
\debut
Res(lhs)\Big \vert_{qA_n+B_n=0} &=&
\({\sig_k^{(n-1)}(B)+B_n\sig_{k-1}^{(n-1)}(B) }\)\non\\
& &~~~~~~~
\times\({ (-q)^{n-1}\prod_{j\not =n}\({\frac{A_j-B_n}{qA_j+B_n}}\)
\cdot\CT^{(n-1)}(P^{(n-1)}|A)}\) Res(P^{(n)}_n) \non
\fin
Similarly applying the recursion relation to the rhs, but
distinguishing whether $A_n$ is a marked point
in the sum (\ref{tqkbis}) or not, we get:
\debut
Res(rhs)\Big \vert_{qA_n+B_n=0}=
&& \left[{(-q)^{n-1}\prod_{j\not =n}\({\frac{A_j-B_n}{qA_j+B_n}}\)\cdot
\CT_k^{(n-1)}(P^{(n-1)}|A) }\right.\non\\
&+& \left.{ (-q)^{n-1} B_n \prod_{j\not =n}\({\frac{A_j-B_n}{qA_j+B_n}}\)
\CT_{k-1}^{(n-1)}(P^{(n-1)}|A) }\right]~Res(P^{(n)}_n) \non
\fin
Comparing these formula gives the equality of the residues
by the induction hypothesis.

ii) Consider the residues at $A_n=A_{n-1}$.
They are both vanishing, and therefore equal.
The proof of the vanishing of the rhs residue is similar to
the proof of vanishing of the lhs residue, so we shall only
present the latter. The potential pole at $A_n=A_{n-1}$
is associated to $\ga_{n,n-1}$. Therefore, its residue may
be computed by using twice the recursion relation (\ref{recurq}).
We get:
\debut
Res(lhs)\Big\vert_{A_n=A_{n-1}}= const.\({
\CT^{(n-2)}(\frac{\ga_{jn}}{\bar \ga_{j,n-1}}P_j|A)P_{n-1}
-\CT^{(n-2)}(\frac{\ga_{j,n-1}}{\bar \ga_{jn}}P_j|A)P_n
}\)\Big\vert_{A_n=A_{n-1}} \non
\fin
This vanishes since we have:
\debut
P_n^{(n)}\Big\vert_{A_n=A_{n-1}}=P_{n-1}^{(n)}\Big\vert_{A_n=A_{n-1}}
\non\fin

iii) Consider now the value of the functions at the point $A_n=B_n$.
At this point we have $P^{(n)}_n=0$ and:
\debut
\ga_{jn} P^{(n)}_j\Big\vert_{A_n=B_n} = P^{(n-1)}_j, \qquad for
\qquad j=1,\cdots,n-1
\non
\fin
Therefore, the recursion relation (\ref{recurq}) yields to:
\debut
lhs\Big\vert_{A_n=B_n} =
\({\sig_k^{(n-1)}(B)+B_n\sig_{k-1}^{(n-1)}(B) }\)~
\CT^{(n-1)}(P^{(n-1)}|A) \non
\fin for the left hand side and,
\debut
rhs\Big\vert_{A_n=B_n} =
\CT_k^{(n-1)}(P^{(n-1)}|A) + B_n \CT_{k-1}^{(n-1)}(P^{(n-1)}|A) \non
\fin
for the right hand side. It clearly appears that they are
equal by the induction hypothesis.

Collecting the points i) to iii) proves the quantum separation
of variables.

The classical formula corresponds to the case $q=1$.

\end{document}